\documentclass[prl, twocolumn,reprint, aps]{revtex4-1}

\usepackage{amsmath,amsfonts, amssymb, amsthm, braket, dsfont}
\usepackage{bm}
\usepackage{mathrsfs}
\usepackage{graphicx}
\usepackage{verbatim}
\usepackage{color}
\usepackage{subfigure}
\usepackage{hyperref}
\usepackage{xr}
\usepackage{pdfpages}

\begin{document}
\title{Symmetry fractionalization of visons in $\mathbb Z_2$ spin
  liquids}
\author{Yang Qi}
\affiliation{Institute for Advanced Study, Tsinghua University,
  Beijing 100084, China}
\affiliation{Perimeter Institute for Theoretical Physics, Waterloo, ON
  N2L 2Y5, Canada} 
\author{Meng Cheng}
\affiliation{Station Q, Microsoft Research, Santa Barbara, California 93106-6105, USA}
\author{Chen Fang}
\affiliation{Department of Physics, Massachusetts Institute of
  Technology, Cambridge, MA 02139, USA}

\begin{abstract}
  In this work we study symmetry fractionalization of vison
  excitations in topological $\mathbb{Z}_2$ spin liquids. We show that
  in the presence of the full $\mathrm{SO}(3)$ spin rotational symmetry and if there is an odd number of spin-$\frac12$
  per unit cell, the symmetry fractionalization of visons is
  completely fixed. On the other hand, visons can have different
  classes of symmetry fractionalization if the spin rotational
  symmetry is reduced.  As a concrete example, we show that visons in
  the Balents-Fisher-Girvin $\mathbb{Z}_2$ spin liquid have crystal
  symmetry fractionalization classes which are not allowed in
  $\mathrm{SO}(3)$ symmetric spin liquids, due to the reduced spin
  rotational symmetry.
\end{abstract} 

\maketitle

Global symmetries, including the spin rotational symmetry, the
time-reversal symmetry and the crystal symmetries, play important
roles in the study of topological spin liquids~\cite{AndersonQSL,  KivelsonPRB1987,
  WenTO1990, Wen1991a, WenZ2SL1991, BalentsSLReview}. Not only are they useful in no-go
theorems that guarantee the existence of topological orders and
constrain their properties when certain symmetries are
unbroken~\cite{OshikawaLSM, HastingsLSM, ZaletelPRL2015, Cheng_unpub}, but they
also further enrich~\cite{XChenLUT} the intrinsic topological orders
of the spin liquids~\cite{WenTO1990, Wen1991a, WenZ2SL1991}, which are
characterized by anyon excitations with fractional braiding
statistics. Particularly, in the resulting symmetry enriched
topological (SET) order the anyon excitations can exhibit symmetry
fractionalization, where the anyons carry fractional quantum numbers of 
the symmetry group~\cite{Essin2013, BarkeshliX, Fidkowski_unpub, Tarantino_arxiv, Teo2015, Lu2013, HungPRB2013}. Examples of symmetry fractionalization include fractionally charged quasiholes
in fractional quantum Hall states 
~\cite{LaughlinFQHE}, and spinons carrying a half-integer spin in
spin liquids~\cite{KalmeyerCSL1987, KivelsonPRB1987}.

Symmetry fractionalization is crucial in classifying, detecting and
modeling the topological order of $\mathbb Z_2$ spin liquids. First,
it refines the classification of topological orders in spin liquids:
although the anyon content is identical, different patterns of symmetry fractionalization
belong to different quantum phases~\cite{wenpsg, YaoFuQi2010X, Essin2013}. Second,
fractional quantum numbers carried by anyons can provide smoking-gun
experimental signatures to detect the anyons and the topological
orders~\cite{Picciotto1997, Saminadayar1997, Han2012}. Lastly, 
symmetry fractionalization plays a key role in the construction of
mean field~\cite{wenpsg, wang2007} and variational wave
functions~\cite{wenpsg, SJiangTPS2015X} for symmetric spin liquids.

In a $\mathbb Z_2$ spin liquid, symmetry fractionalization is fully
characterlized by the symmetry quantum numbers carried by spinons
(either the bosonic or fermionic ones) and visons,
respectively~\cite{Essin2013}. The symmetry fractionalization of
spinons has been extensively studied using parton
constructions~\cite{wenpsg, wang2007, YMLuKagomePSG2011, LuBFU,
  Zheng_arxiv, YMLuTri2015X, FWangSq2015X} and tensor network
states~\cite{SJiangTPS2015X}. However in all these studies it turns
out that the symmetry fractionalization of visons is completely determined by the background spinons on the lattice. Recently it
has been shown that certain SET phases where spinons and visons both
have nontrivial symmetry fractionalizations are
anomalous~\cite{VishwanathPRX2013, CWangETMT2013, Qi_unpub,
  HermeleFFAT}, implying that these phases cannot possibly exist in any
two-dimensional (2D) systems and in fact can only be realized on the
surface of  three-dimensional (3D) symmetry protected topological
(SPT) phases~\cite{XChenSPT}, such as 3D topological
insulator/superconductors (TSC)~\cite{SchnyderTSC2008} and 
topological cystalline insulators (TCI)~\cite{FuTCI2011}. However,
although there is a general framework to detect anomalous SET orders
for onsite unitary symmetries~\cite{BarkeshliX, Chen2014}, for SET
orders with time-reversal symmety and crystal symmetries a general
method of detecting anomaly is still lacking. Since evidences for
gapped $\mathbb Z_2$ spin liquid phases with full spin rotational, time
reversal and crystal symmetries have been found in numerical studies
of spin-$\frac12$ Heisenberg models on the kagome 
lattices~\cite{YanScience, JiangNatPhys, DepenbrockPRL2012} (see
however Ref.~\onlinecite{IqbalKagome2011}), and also in a recent experiment in
the spin-$\frac12$ kagome antiferromagnet
herbertsmithite~\cite{YSLeeNMR_aps}, it is an important theoretical question to study the
classification of (anomaly-free) symmetry fractionalization in fully
symmetric gapped $\mathbb Z_2$ spin liquids.

In this work we study symmetry fractionalization for vison excitations
in a $\mathbb Z_2$ spin liquid with spin rotational, time-reversal and
crystal symmetries. Generalizing an elegant method to detect anomalies recently proposed by~\citet{HermeleFFAT}, we show that on kagome or square lattices, with full SO(3) spin rotational
symmetry and under the assumption that the spinon carries a half-integer
spin, all symmetry fractionalization classes for visons are uniquely fixed,
except the commutation relation between the two unit translations.
 In fact, any SET orders where visons exhibit other nontrivial
symmetry fractionalization are anomalous. 
Our conclusion crucially depends on
the full SO(3) spin rotational symmetry, and we demonstrate using the Balents-Fisher-Girvin(BFG) model~\cite{BFGZ2SL2002} that the visons can exhibit more complicated fractionalization patterns once the spin rotational symmetry is reduced to $\mathrm{O}(2)$.

\emph{Symmetry fractionalization of vison in $\mathrm{SO}(3)$
  symmetric spin liquids.} We consider $\mathbb Z_2$ spin liquids enriched by the
symmetry group
$\mathrm{SO}(3)\times \mathbb{Z}_2^T\times G_{\text{space}}$, denoting
the spin rotational symmetry, the time-reversal symmetry and space
group of the lattice, respectively. We will focus on the 
kagome lattice and treat the more complicated case of square lattice
in Sec.~\ref{sec:square} of the Supplemental Material. Notice that for systems consist of half-integer spins, the symmetry group should still be considered $\mathrm{SO}(3)$ instead of $\mathrm{SU}(2)$ because local excitations, such as the magnons, always carry integer spins. We will also assume that there are even number of sites in the system, so the ground state can be a spin singlet.

Before we discuss the
fractionalization of space group and time-reversal symmetries, we
first determine the $\mathrm{SO}(3)$ spin quantum numbers of the
anyons. A $\mathbb{Z}_2$ spin liquid has four types of topologically distinct quasiparticle excitations: the trivial excitation $\mathds{1}$, the bosonic spinon $e$ and the vison $m$ which are mutually semionic, and the fermionic spinon $\epsilon=e\times m$. Here we assume that the $e$ anyon in the $\mathbb Z_2$ spin
liquid carries a half-integer spin projective representation of the
SO(3) symmetry group. With this assumption one can further show that the
time-reversal symmetry forces the vison $m$ to have an integer spin,
otherwise the system must have a nontrivial Hall response associated with
the $S^z$ charge, which necessarily breaks the time-reversal symmetry~\cite{VishwanathPRX2013, metlitski2013}.
On a lattice with an odd number of spin-$\frac12$ per unit cell (which 
includes most of the gapped spin liquids found in numerical simulations so far~\cite{YanScience, JiangNatPhys, DepenbrockPRL2012, ZhuPRB2015,  WJHuTriZ2SL2015, TayVMC2011, Rousochatzakis2014}, and the Lieb-Schultz-Mattis-Oshikawa-Hastings theorem~\cite{LSM, OshikawaLSM, HastingsLSM} guarantees that any gapped symmetric states found in this sort of system must be topologically ordered),  it can be
shown that~\cite{ZaletelPRL2015, Cheng_unpub}: (a) our assumption of $e$
carrying a half-integer spin is automatically guaranteed, and (b) the
vison sees a $\pi$-flux when moving around a unit cell, i. e. the
vison transforms projectively under translation symmetries:
$T_1T_2=-T_2T_1$ where $T_{1,2}$ are translations by the two basis vectors of the Bravais lattice, because when vison moves around a unit cell it
braids with an odd number of spinons inside the unit cell, and each braiding gives a
$-1$ Berry phase.

\begin{figure}[t!]
  \centering
  \subfigure[\label{fig:fluxes:au}]{\includegraphics{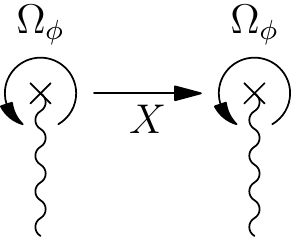}}\qquad\qquad
  \subfigure[\label{fig:fluxes:u}]{\includegraphics{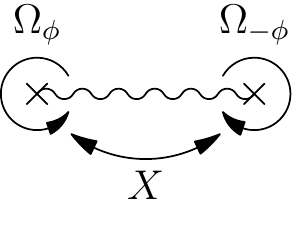}}
  \caption{Symmetry actions on fluxes. The cross labels the position
    of fluxes, and the wiggly lines mark the positions of the
    branch-cuts. (a) An antiunitary symmetry $X$ maps a flux to
    itself. (b) A unitary symmetry $X$ maps two fluxes with opposite
    $\phi$ to each other.}
  \label{fig:fluxes}
\end{figure}

Our argument of fixing vison's symmetry fractionalization is based on
the flux-fusion anomaly test recently proposed
by~\citet{HermeleFFAT}. We will consider systems on a disk or an infinite plane. In this test, we adiabatically insert fluxes of the U(1)
global symmetry of spin rotations with respect to $S^z$. A flux
$\Omega_\phi$ can be inserted by applying the unitary
transformation $U=e^{\frac{i\phi}{2\pi}\sum_\mathbf{r}\theta_\mathbf{r}S_\mathbf{r}^z}$, where $\mathbf{r}$ labels lattice sites and $\theta_\mathbf{r}$
denotes the polar angle of $\mathbf{r}$ with respect to a branch cut, as shown in
Fig.~\ref{fig:fluxes}. Because $e$ and $\epsilon$ anyons carry
half-integer spins, a $2\pi$ flux carries the topological charge of a vison,
$\Omega_{2\pi}\sim m$~\cite{BarkeshliX, HermeleFFAT}. If the symmetry fractionalization of $m$ can be deduced from a physical invariant of $2\pi$ fluxes, which can be extended to a well-defined quantity for the whole adiabatic flux-insertion process, based on continuity we can immediately relate the symmetry
fractionalization of $\Omega_{2\pi}\sim m$ to that of the vacuum sector $\mathds{1}$, which is always trivial.~\footnote{We notice that the adiabatic insertion is not essential for the flux-fusion argument to work, and it can be generalized to a discrete subgroup of the $\mathrm{U}(1)$ symmetry, see Ref. [\onlinecite{HermeleFFAT}].} 

We will now expand on this general method, focusing on an order $2$ symmetry operation $X$. As we will see later this is sufficient for our purpose. 

For an anti-unitary $\mathbb Z_2$ symmetry operation $X$, we implement
$X$ such that a flux $\Omega_\phi$ at position $\mathbf{r}$ is mapped to
itself under $X$:
\begin{equation}
  \label{eq:Xau}
  X:\Omega_\phi(\mathbf{r})\rightarrow \Omega_\phi(\mathbf{r}),
\end{equation}
as shown in Fig.~\ref{fig:fluxes:au}, and thus one can define the
local action of $X$. Two topologically distinct ways $X$ can act are
given by $X^2=\pm 1$ on the flux, corresponding to the flux being a
Kramers singlet/doublet under $X$. For $0<\phi<2\pi$, mathematically
this is the well-known fact that the second group cohomology with $\mathrm{U}(1)$ coefficient
$H^2(\mathbb{Z}_2^X, \mathrm{U}(1))=\mathbb{Z}_2$ ( $\mathbb Z_2^X$
denotes the group generated by $X$), which classifies projective representations of $\mathbb{Z}_2^X$~\cite{XChenSPT}. In the end of flux insertion when
a vison is created, due to the $\mathbb{Z}_2$ fusion rule
$m\times m=\mathds{1}$ the local $X$ action is classified by
$H^2(\mathbb{Z}_2^X, \mathbb{Z}_2)=\mathbb{Z}_2$~\cite{Essin2013,
  BarkeshliX}, distinguished again by the local $X^2$ value.
Therefore the two classifications coincide, and the local $X^2$ value
is well defined for the whole flux-insertion process. We therefore
conclude that $m$ must have $X^2=+1$.

For an unitary $\mathbb Z_2$ space group symmetry operation $X$, we can not use the same argument when the flux is mapped to itself under $X$. The reason is that for $0<\phi<2\pi$, the local value of $X^2$ is no longer a physical invariant because one can redefine the local action of $X$ by an arbitrary $\mathrm{U}(1)$ phase, consistent with $H^2(\mathbb Z_2^X, \mathrm{U}(1))=\mathbb Z_1$. But for a vison, $X^2=\pm 1$ are still topologically distinct due to the constraint from the $\mathbb{Z}_2$ fusion rule in accordance with $H^2(\mathbb{Z}^X_2, \mathbb{Z}_2)=\mathbb{Z}_2$.
Therefore for an unitary symmetry operation we need to consider
a different setup where $X$ maps a flux $\Omega_\phi$ located at
position $\mathbf{r}$ to a flux $\Omega_{-\phi}$ at a symmetry-related
position $ X\mathbf{r}$, as shown in Fig.~\ref{fig:fluxes:u}.
\begin{equation}
  \label{eq:Xu}
  X:\Omega_\phi(\mathbf{r})\rightarrow\Omega_{-\phi}(X\mathbf{r}).
\end{equation}
Now we consider a configuration with two fluxes inserted:
$\Omega_\phi(\mathbf{r})\Omega_{-\phi}(X\mathbf{r})$. Since the total
flux is zero, this configuration can be consistently put on a finite
system with a $\phi$-independent open boundary conditions and has a
well-defined $X$-symmetry parity eigenvalue $\lambda_X(\phi)=\pm1$. In
the limit of $\phi\rightarrow2\pi$, the two fluxes become two visons,
and the ratio of parity eigenvalues $\lambda_X(2\pi)/\lambda_X(0)$
gives the symmetry fractionalization of vison~\cite{LuBFU, QiCSF,
  ZLVPSG}. If the symmetry $X$ is unbroken for any $\phi$, the parity
eigenvalue $\lambda_X(\phi)$ cannot jump between $+1$ and $-1$, and
therefore the vison must carry $X^2=+1$.

In summary, using the flux-fusion anomaly test we can conclude that
the vison can only take a trivial symmetry fractionalization of
$X^2=+1$ if the flux $\Omega_\phi$ transforms under $X$ as described
by Eq.~\eqref{eq:Xau} if $X$ is anti-unitary, and Eq.~\eqref{eq:Xu} if
$X$ is unitary.

\begin{figure}[t!]
  \centering
  \subfigure[\label{fig:symop}]{\includegraphics{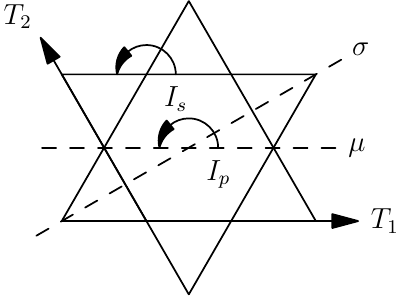}}
  \subfigure[\label{fig:dual}]{\includegraphics{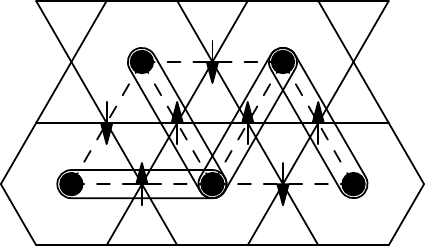}}
  \caption{(a) Symmetry operations on a kagome lattice. $T_1$ and
    $T_2$ denote translations by the two basis vectors of the Bravais
    lattice; $\mu$ and $\sigma$ denote mirror reflections with respect
    to the axes shown by the dashed line; $I_p$ and $I_s$ denote
    plaquette-centered and site-centered inversions (i.e. 180-degree
    rotation), respectively. (b) Duality mapping from the BFG
    model~\cite{BFGZ2SL2002} to the quantum dimer model. The spins on
    the kagome lattice sites are mapped to dimers on the bonds of a
    dual triangular lattice: $S^z=+\frac12$ ($-\frac12$) maps to dimer
    (no dimer), respectively.}
  \label{fig:kagome}
\end{figure}

These criteria significantly constrain the possible symmetry
fractionalizations of the vison, in the presence of $\mathrm{U}(1)$
spin rotational symmetry. However they are not sufficient to fix all
the vison fractionalization classes. To be concrete, we consider the
example of the kagome lattice, where the symmetry group is
$G=\mathrm{SO}(3)\times p6m\times \mathbb Z_2^T$, with $p6m$ being the
space symmetry group of the kagome lattice and $\mathbb Z_2^T$ the
group generated by time-reversal symmetry. The fractionalization of
this symmetry group on visons are classified by
$H^2(G, \mathbb Z_2)=\mathbb Z_2^7$, and are labeled by seven
$\mathbb Z_2$ invariants: $\omega_{12}$, $\omega_\mu$,
$\omega_\sigma$, $\omega_I$, $\omega_T$, $\omega_{\mu T}$ and
$\omega_{\sigma T}$, as listed in Table~\ref{tab:psg}. Here $\mu$,
$\sigma$ denote two mirror reflections and $I_p=(\mu\sigma)^3$ denotes
the plaquette-centered inversion, as shown in
Fig.~\ref{fig:symop}. The first variable $\omega_{12}$ labels the
commutation relation fractionalization~\cite{QiCSF}
$T_1T_2=\pm T_2T_1$ and the other six variables, in the form of
$\omega_X$, label quantum number fractionalization $X^2=\pm1$.

\begin{table}[htbp]
  \centering
  \caption{Quantum numbers labeling different symmetry
    fractionalization of visons. The first column lists the labels
    indicating fractional quantum numbers associated with the
    algebraic relations listed in the second column. The third column
    labels symmetry operation $X^\prime$ used in the anomaly test, and
    the last column lists the symmetry required to derive $X^2=+1$.}
  \begin{tabular*}{\columnwidth}{@{\extracolsep{\fill}}cccc}
    \hline\hline
    Label & Algebraic relation & $X^\prime$
    & Required symmetry\\
    \hline
    $\omega_{12}$ & $T_1T_2=\pm T_2T_1$ & -- & --\\
    $\omega_\mu$ & $\mu^2=\pm1$ & $\mu$ & U(1)\\
    $\omega_\sigma$ & $\sigma^2=\pm1$ & $\sigma$ & U(1)\\
    $\omega_T$ & $T^2=\pm1$ & $T$ & U(1)\\
	$\omega_I$ & $I_p^2=\pm1$ & $I_pe^{i\pi S^y}$ & $\mathrm{SO}(3)$\\
	$\omega_{\mu T}$ & $(\mu T)^2=\pm1$ & $\mu Te^{i\pi S^y}$ & $\mathrm{SO}(3)$\\
    $\omega_{\sigma T}$ & $(\sigma T)^2=\pm1$ & $\sigma Te^{i\pi S^y}$ 
	& $\mathrm{SO}(3)$\\
    \hline\hline
  \end{tabular*}
  \label{tab:psg}
\end{table}

Next we divide the symmetry operations $X$  into two
classes according to how they act on the aforementioned
$\mathrm{U}(1)_{S_z}$ flux $\Omega_\phi$: the symmetry operations $T$,
$\mu$ and $\sigma$ satisfy either Eq.~\eqref{eq:Xau} or
Eq.~\eqref{eq:Xu} (depending on whether the operation is unitary or
anti-unitary) and can be used directly in the anomaly test, which
immediately gives the constraints that vison must carry trivial
symmetry fractionalization $\omega_T=\omega_\mu=\omega_\sigma=1$. We
note that these symmetries satisfy $t(X)=-1$ for the function $t$
defined in Ref.~\onlinecite{HermeleFFAT}, and the constraints $X^2=+1$
have all been obtained in previous works~\cite{CWangETMT2013, Qi_unpub,
 HermeleFFAT}.

The other three symmetry operations $I_p$, $\mu T$ and $\sigma T$ do
not satisfy either condition and cannot be used directly in the
anomaly test [they satisfy $t(X)=1$].  To make progress, we generalize
the flux-fusion argument by including spin rotations: we observe that
if we combine $\mu T$ or $\sigma T$ with a spin rotation
$e^{i\pi S^y}$ which reverses the symmetry flux $\phi$, the new
symmetry operations now preserve the $\mathrm{U}(1)$ fluxes and
therefore satisfy Eq.~\eqref{eq:Xau}. Similarly, $I_pe^{i\pi S^y}$
satisfies Eq.~\eqref{eq:Xu}. Hence the combined symmetry
$X^\prime=Xe^{i\pi S^y}$ can be used in the anomaly test to infer that
the vison must have $(Xe^{i\pi S^y})^2=+1$ for $X=I_p, \mu T$ and
$\sigma T$. We also know that the vison carries $(e^{i\pi S^y})^2=+1$
because it has an integer spin, and one can further show due to the
connectedness of $\mathrm{SO}(3)$ group that the spin rotation
$e^{i\pi S^y}$ must commute with $X$ (see Sec.~\ref{sec:fract-sc} of
the Supplemental Material
for more details).  Therefore from the identity
$(Xe^{i\pi S^y})^2=X^2(e^{i\pi S^y})^2 Xe^{i\pi S^y}X^{-1}e^{-i\pi
  S^y}$
we find that $X^2=+1$ on visons for $X=I_p$, $\mu T$ and $\sigma T$.

To summarize, using the anomaly test we show that in a SO(3) symmetric
$\mathbb Z_2$ spin liquid where the spinons carry a half-integer spin,
the vison must carry trivial quantum number fractionalization except
for $\omega_{12}$ as listed in Table~\ref{tab:psg}. In fact, the
symmetry fractionalization of the visons is completely captured by an
Ising gauge theory. [The Ising gauge theory is even (odd) if there are
an even (odd) number of spin-$\frac12$ per unit cell, respectively.]
As listed in Table~\ref{tab:psg}, deriving $\omega_X=+1$ for
$X=T,\sigma,\mu$ only requires a U(1) subgroup of the spin SO(3)
symmetry, but deriving $\omega_X=+1$ for the other three
$X=I_p, \sigma T, \mu T$ requires the full SO(3) symmetry. In
particular, although in our argument we only use the symmetry
operations from the $\mathrm U(1)\rtimes\mathbb Z_2=\mathrm O(2)$
subgroup generated by $S^z$ and $e^{i\pi S^y}$, the connetedness of
SO(3) plays a crucial role in enforcing that $e^{i\pi S^y}$ commutes
with $X$ in the fractionalization class. As we will see
below, the vison can exhibit more complicated symmetry
fractionalizations if the spin rotational symmetry is reduced to O(2)
or a discrete subgroup. More details can be found in
Sec.~\ref{sec:fract-sc} of the Supplemental Material.

We note that when introducing the fluxes, a branch-cut as shown in
Fig.~\ref{fig:fluxes} is also introduced. To consistently define the
symmetry transformations of the fluxes, this branch-cut should be
chosen to be symmetric under $X$. However, such a choice is impossible
for a site-centered inversion symmetry $I_s$ [see
Fig.~\ref{fig:symop}], and therefore our argument cannot be applied
directly to fix $I_s^2=+1$. This is discussed in more details in
Sec.~\ref{sec:site-cent-invers} of the Supplemental Material.

\emph{$\mathbb Z_2$ SET with nontrivial vison symmetry
  fractionalization.} Our argument showing the vison cannot have
nontrivial symmetry fractionalization requires the spin liquid to have
the full spin SO(3) symmetry. Hence breaking the spin rotational
symmetry down to U(1) or $\mathrm O(2)=\mathrm U(1)\rtimes\mathbb Z_2$
opens up the possibility for the visons to have nontrivial
fractionalized quantum numbers of the symmetries
$X=I_p, \mu T, \sigma T$, as listed in Table~\ref{tab:psg}. Here we
present a concrete example of $\mathbb Z_2$ SET phases, where the $e$
anyon has half $\mathrm{U}(1)$ charge and the vison $m$ has 
both $I_p^2=-1$ and $(\mu T)^2=-1$.

We consider the $\mathbb Z_2$ SET phase found in the
BFG spin-$1/2$ XXZ model on the kagome
lattice~\cite{BFGZ2SL2002}. In this model the spin rotational symmetry
is broken from SO(3) down to O(2) due to the easy-axis anisotropy.
Therefore our argument for vison having
$I_p^2=(\mu T)^2=(\sigma T)^2=+1$ no longer applies, although we can
still use the anomaly test to show
$(I_pe^{i\pi S^y})^2=(\mu Te^{i\pi S^y})^2=(\sigma Te^{i\pi
  S^y})^2=+1$.
Indeed the vison in this $\mathbb Z_2$ phase carries $I_p^2=-1$ and
$(\mu T)^2=-1$.

To see that the vison carries $I_p^2=-1$, we notice that the
Hamiltonian can be mapped to a quantum dimer model (QDM) with three
dimers per site on a triangular lattice~\cite{BFGZ2SL2002}, where each spin, viewed as a
hard-core boson, corresponds to a dimer in the latter model, as shown
in Fig.~\ref{fig:dual}. In this mapping the sites of the kagome
lattice map to the bonds of the dual triangular lattice, and the sites
of the dual triangular lattice map to the center of the hexagons of
the kagome lattice. Therefore the plaquette-centered inversion $I_p$
on the kagome lattice [see Fig.~\ref{fig:symop}] becomes a
site-centered inversion $I_s^\text{dual}$ on the triangular
lattice. It is well-known that in the $\mathbb Z_2$ topological phase
of a quantum dimer model with an odd number of dimers per site, the
vison excitations are described by an odd Ising gauge
theory~\cite{SachdevFFIM1999, SenthilZ2SL2000, MoessnerTri2001, MoessnerZ2Gauge2001},
which implies that they carry a nontrivial symmetry fractionalization
$(I_s^\text{dual})^2=-1$ for the site-centered inversion of the
triangular lattice~\cite{Zheng_arxiv, YMLuTri2015X}. Therefore the vison excitations
in this $\mathbb Z_2$ spin liquid have $I_p^2=-1$. We can also say
that the background spinons are now located at the centers of the
hexagons although the physical spin-$1/2$'s are at sites, which is
actually evident from the mapping to the quantum dimer model, and the
symmetry fractionalization $I_p^2=-1$ comes from the braiding between
the visons and the spinon at the inversion center when applying $I_p$
to the configuration in Fig.~\ref{fig:fluxes:u}.

In fact, our generalized flux-fusion argument is circumvented
precisely by the anticommutation between $e^{i\pi S^y}$ and $I_p$ on
visons, which is allowed for O(2) spin symmetry but not for SO(3). As
shown in Sec.~\ref{sec:bfg} of the Supplemental Material in the fractionalization class of the
vison the spin rotation $e^{i\pi S^y}$ commutes with $\sigma$ but
anticommutes with $\mu$, so it anticommutes with
$I_p=(\mu\sigma)^3$. Therefore combined with the fact that
$(e^{i\pi S^y})^2=+1$ on the vison, the anomaly test gives
$(I_pe^{i\pi S_y})^2=-I_p^2 (e^{i\pi S_y})^2$ which implies
$I_p^2=-1$.

Similarly, using the flux-fusion anomaly test we have $(\mu T e^{i\pi S^y})^2=+1$.  However as we
show in Sec.~\ref{sec:bfg} of the Supplemental Material, when acting on the vison $e^{i\pi S^y}$ commutes with $T$ but anticommutes with
$\mu$, and therefore it anticommutes with $\mu T$. Hence we get
$(\mu T)^2=-1$ for visons.

These results have interesting physical implications. First of all, since in a $\mathrm{SO}(3)$-symmetric spin liquid visons have $I_p^2=(\mu T)^2=+1$, we  have shown that the $\mathbb Z_2$ spin liquid in the BFG
model cannot be smoothly connected to the $\mathbb Z_2$ spin liquid
state found in the antiferromagnetic Heisenberg model on the kagome
lattice with full SO(3) spin rotational symmetry, without breaking the space symmetry~\footnote{To better
  compare the two phases we can perturb the antiferomagnetic
  Heisenberg model slightly by introducing easy-axis anisotropy. Such
  a nearest-neighbor XXZ model was studied numerically in
  Ref. \onlinecite{HePRL2015}, and it was established that the
  $\mathbb{Z}_2$ spin liquid phase persists when the easy-axis
  anisotropy is turned on. Ref. \onlinecite{HePRL2015} also made the
  observation that the BFG model and the Heisenberg model describe
  different $\mathbb{Z}_2$ spin liquids, but the claimed reason that
  the spinons in the BFG model are Kramers singlet is incorrect. }. Secondly, the nontrivial fractionalization class $(\mu T)^2=-1$ for visons implies the existence of gapless edge states protected by both $T$ and $\mu$, on an edge that preserves the reflection $\mu$~\cite{LuBFU}.

\emph{Conclusion.} In this work we show that in a $\mathbb Z_2$ spin
liquid with an unbroken SO(3) spin rotational symmetry, the symmetry
fractionalization of visons is completely fixed if the spinon carries
a half-integer spin. The last condition is automatically fulfilled in
many candidate systems of $\mathbb Z_2$ spin liquids studied so
far~\cite{YanScience, JiangNatPhys, DepenbrockPRL2012, ZhuPRB2015,
  WJHuTriZ2SL2015} as they have an odd number of spin-$\frac12$ per
unit cell~\cite{ZaletelPRL2015, Cheng_unpub}. Our result can be
directly applied to simplify the numerical detection~\cite{ZLVPSG,
  QiCSF} of the symmetry fractionalization in $\mathbb Z_2$ spin
liquids, as only the fractional quantum numbers of one type of spinons
(either $e$ or $\epsilon$) need to be measured, and those of $m$ and
the other spinon can then be determined theoretically~\cite{LuBFU,
  Zheng_arxiv, YMLuTri2015X}.

Having determined the symmetry fractionalization of visons, one
still has a wide variety of possible $\mathbb{Z}_2$ spin liquids with
different symmetry fractionalization classes of spinons. Representative wave
functions for spin liquids with different symmetry fractionalizations
are constructed using parton constructions including Schwinger boson
and Abrikosov fermions~\cite{sstri, wang2007, YMLuKagomePSG2011,
  LuBFU, Zheng_arxiv, YMLuTri2015X, FWangSq2015X}, as well as tensor
network states~\cite{SJiangTPS2015X}. However, in all these
constructions some spinon symmetry fractionalization classes can only be
realized as \emph{gapless} $\mathbb Z_2$ spin liquids, and we will
leave the question of existence of gapped wavefunctions for those classes to future works.

Our argument suggests that to look for spin liquids with nontrivial
vison symmetry fractionalization, one should turn to systems with
reduced spin rotational symmetry. We consider the $\mathbb{Z}_2$ spin liquid in the BFG model as an example,  where the spin rotational
symmetry is reduced to O(2). We find that quite interestingly the visons carry $I_p^2=-1$ as well as $(\mu T)^2=-1$. This example only scratches the surface of the
rich possibilities of different SET orders in models with reduced spin
rotational symmetry, which can be realized in systems where strong
electron interactions interplay with strong spin-orbit couplings. We
leave the classification and construction of symmetric spin liquids
in these systems to future works.

\begin{acknowledgements}
  We thank Bela Bauer, Lukasz Cincio, Liang Fu and Zheng-Cheng Gu for
  enlightening discussions, and in particular Michael Hermele and Xie
  Chen for comments on the manuscript and Yuan-Ming Lu for helpful
  correspondence on topological superconductors.  Y.Q. is supported by
  National Basic Research Program of China through Grant
  No. 2011CBA00108. CF is supported by S3TEC Solid State Solar Thermal
  Energy Conversion Center, an Energy Frontier Research Center funded
  by the U.S. Department of Energy (DOE), Office of Science, Basic
  Energy Sciences (BES), under Award
  No. de-sc0001299/DE-FG02-09ER46577. This research was supported in
  part by Perimeter Institute for Theoretical Physics. Research at
  Perimeter Institute is supported by the Government of Canada through
  Industry Canada and by the Province of Ontario through the Ministry
  of Research and Innovation.
\end{acknowledgements}

\bibliography{psg}

\begin{thebibliography}{62}%
\makeatletter
\providecommand \@ifxundefined [1]{%
 \@ifx{#1\undefined}
}%
\providecommand \@ifnum [1]{%
 \ifnum #1\expandafter \@firstoftwo
 \else \expandafter \@secondoftwo
 \fi
}%
\providecommand \@ifx [1]{%
 \ifx #1\expandafter \@firstoftwo
 \else \expandafter \@secondoftwo
 \fi
}%
\providecommand \natexlab [1]{#1}%
\providecommand \enquote  [1]{``#1''}%
\providecommand \bibnamefont  [1]{#1}%
\providecommand \bibfnamefont [1]{#1}%
\providecommand \citenamefont [1]{#1}%
\providecommand \href@noop [0]{\@secondoftwo}%
\providecommand \href [0]{\begingroup \@sanitize@url \@href}%
\providecommand \@href[1]{\@@startlink{#1}\@@href}%
\providecommand \@@href[1]{\endgroup#1\@@endlink}%
\providecommand \@sanitize@url [0]{\catcode `\\12\catcode `\$12\catcode
  `\&12\catcode `\#12\catcode `\^12\catcode `\_12\catcode `\%12\relax}%
\providecommand \@@startlink[1]{}%
\providecommand \@@endlink[0]{}%
\providecommand \url  [0]{\begingroup\@sanitize@url \@url }%
\providecommand \@url [1]{\endgroup\@href {#1}{\urlprefix }}%
\providecommand \urlprefix  [0]{URL }%
\providecommand \Eprint [0]{\href }%
\providecommand \doibase [0]{http://dx.doi.org/}%
\providecommand \selectlanguage [0]{\@gobble}%
\providecommand \bibinfo  [0]{\@secondoftwo}%
\providecommand \bibfield  [0]{\@secondoftwo}%
\providecommand \translation [1]{[#1]}%
\providecommand \BibitemOpen [0]{}%
\providecommand \bibitemStop [0]{}%
\providecommand \bibitemNoStop [0]{.\EOS\space}%
\providecommand \EOS [0]{\spacefactor3000\relax}%
\providecommand \BibitemShut  [1]{\csname bibitem#1\endcsname}%
\let\auto@bib@innerbib\@empty
\bibitem [{\citenamefont {Anderson}(1973)}]{AndersonQSL}%
  \BibitemOpen
  \bibfield  {author} {\bibinfo {author} {\bibfnamefont {P.}~\bibnamefont
  {Anderson}},\ }\href@noop {} {\bibfield  {journal} {\bibinfo  {journal}
  {Mater. Res. Bull.}\ }\textbf {\bibinfo {volume} {8}},\ \bibinfo {pages}
  {153} (\bibinfo {year} {1973})}\BibitemShut {NoStop}%
\bibitem [{\citenamefont {Kivelson}\ \emph {et~al.}(1987)\citenamefont
  {Kivelson}, \citenamefont {Rokhsar},\ and\ \citenamefont
  {Sethna}}]{KivelsonPRB1987}%
  \BibitemOpen
  \bibfield  {author} {\bibinfo {author} {\bibfnamefont {S.~A.}\ \bibnamefont
  {Kivelson}}, \bibinfo {author} {\bibfnamefont {D.~S.}\ \bibnamefont
  {Rokhsar}}, \ and\ \bibinfo {author} {\bibfnamefont {J.~P.}\ \bibnamefont
  {Sethna}},\ }\href {http://link.aps.org/doi/10.1103/PhysRevB.35.8865}
  {\bibfield  {journal} {\bibinfo  {journal} {Phys. Rev. B}\ }\textbf {\bibinfo
  {volume} {35}},\ \bibinfo {pages} {8865(R)} (\bibinfo {year}
  {1987})}\BibitemShut {NoStop}%
\bibitem [{\citenamefont {Wen}(1990)}]{WenTO1990}%
  \BibitemOpen
  \bibfield  {author} {\bibinfo {author} {\bibfnamefont {X.~G.}\ \bibnamefont
  {Wen}},\ }\href {\doibase 10.1142/s0217979290000139} {\bibfield  {journal}
  {\bibinfo  {journal} {Int. J. Mod. Phys. B}\ }\textbf {\bibinfo {volume}
  {04}},\ \bibinfo {pages} {239} (\bibinfo {year} {1990})}\BibitemShut
  {NoStop}%
\bibitem [{\citenamefont {Wen}(1991{\natexlab{a}})}]{Wen1991a}%
  \BibitemOpen
  \bibfield  {author} {\bibinfo {author} {\bibfnamefont {X.-G.}\ \bibnamefont
  {Wen}},\ }\href {\doibase 10.1142/s0217979291001541} {\bibfield  {journal}
  {\bibinfo  {journal} {Int. J. Mod. Phys. B}\ }\textbf {\bibinfo {volume}
  {05}},\ \bibinfo {pages} {1641} (\bibinfo {year}
  {1991}{\natexlab{a}})}\BibitemShut {NoStop}%
\bibitem [{\citenamefont {Wen}(1991{\natexlab{b}})}]{WenZ2SL1991}%
  \BibitemOpen
  \bibfield  {author} {\bibinfo {author} {\bibfnamefont {X.~G.}\ \bibnamefont
  {Wen}},\ }\href {\doibase 10.1103/PhysRevB.44.2664} {\bibfield  {journal}
  {\bibinfo  {journal} {Phys. Rev. B}\ }\textbf {\bibinfo {volume} {44}},\
  \bibinfo {pages} {2664} (\bibinfo {year} {1991}{\natexlab{b}})}\BibitemShut
  {NoStop}%
\bibitem [{\citenamefont {Balents}(2010)}]{BalentsSLReview}%
  \BibitemOpen
  \bibfield  {author} {\bibinfo {author} {\bibfnamefont {L.}~\bibnamefont
  {Balents}},\ }\href {\doibase 10.1038/nature08917} {\bibfield  {journal}
  {\bibinfo  {journal} {Nature}\ }\textbf {\bibinfo {volume} {464}},\ \bibinfo
  {pages} {199} (\bibinfo {year} {2010})}\BibitemShut {NoStop}%
\bibitem [{\citenamefont {Oshikawa}(2000)}]{OshikawaLSM}%
  \BibitemOpen
  \bibfield  {author} {\bibinfo {author} {\bibfnamefont {M.}~\bibnamefont
  {Oshikawa}},\ }\href@noop {} {\bibfield  {journal} {\bibinfo  {journal}
  {Phys. Rev. Lett.}\ }\textbf {\bibinfo {volume} {84}},\ \bibinfo {pages}
  {1535} (\bibinfo {year} {2000})}\BibitemShut {NoStop}%
\bibitem [{\citenamefont {Hastings}(2004)}]{HastingsLSM}%
  \BibitemOpen
  \bibfield  {author} {\bibinfo {author} {\bibfnamefont {M.~B.}\ \bibnamefont
  {Hastings}},\ }\href@noop {} {\bibfield  {journal} {\bibinfo  {journal}
  {Phys. Rev. B}\ }\textbf {\bibinfo {volume} {69}},\ \bibinfo {pages} {104431}
  (\bibinfo {year} {2004})}\BibitemShut {NoStop}%
\bibitem [{\citenamefont {Zaletel}\ and\ \citenamefont
  {Vishwanath}(2015)}]{ZaletelPRL2015}%
  \BibitemOpen
  \bibfield  {author} {\bibinfo {author} {\bibfnamefont {M.~P.}\ \bibnamefont
  {Zaletel}}\ and\ \bibinfo {author} {\bibfnamefont {A.}~\bibnamefont
  {Vishwanath}},\ }\href@noop {} {\bibfield  {journal} {\bibinfo  {journal}
  {Phys. Rev. Lett.}\ }\textbf {\bibinfo {volume} {114}},\ \bibinfo {pages}
  {077201} (\bibinfo {year} {2015})}\BibitemShut {NoStop}%
\bibitem [{\citenamefont {Cheng}\ \emph {et~al.}()\citenamefont {Cheng},
  \citenamefont {Zaletel}, \citenamefont {Barkeshli}, \citenamefont
  {Bonderson},\ and\ \citenamefont {Vishwanath}}]{Cheng_unpub}%
  \BibitemOpen
  \bibfield  {author} {\bibinfo {author} {\bibfnamefont {M.}~\bibnamefont
  {Cheng}}, \bibinfo {author} {\bibfnamefont {M.~P.}\ \bibnamefont {Zaletel}},
  \bibinfo {author} {\bibfnamefont {M.}~\bibnamefont {Barkeshli}}, \bibinfo
  {author} {\bibfnamefont {P.}~\bibnamefont {Bonderson}}, \ and\ \bibinfo
  {author} {\bibfnamefont {A.}~\bibnamefont {Vishwanath}},\ }\href@noop {}
  {}\bibinfo {note} {To appear}\BibitemShut {NoStop}%
\bibitem [{\citenamefont {Chen}\ \emph {et~al.}(2010)\citenamefont {Chen},
  \citenamefont {Gu},\ and\ \citenamefont {Wen}}]{XChenLUT}%
  \BibitemOpen
  \bibfield  {author} {\bibinfo {author} {\bibfnamefont {X.}~\bibnamefont
  {Chen}}, \bibinfo {author} {\bibfnamefont {Z.-C.}\ \bibnamefont {Gu}}, \ and\
  \bibinfo {author} {\bibfnamefont {X.-G.}\ \bibnamefont {Wen}},\ }\href
  {\doibase 10.1103/PhysRevB.82.155138} {\bibfield  {journal} {\bibinfo
  {journal} {Phys. Rev. B}\ }\textbf {\bibinfo {volume} {82}},\ \bibinfo
  {pages} {155138} (\bibinfo {year} {2010})}\BibitemShut {NoStop}%
\bibitem [{\citenamefont {Essin}\ and\ \citenamefont
  {Hermele}(2013)}]{Essin2013}%
  \BibitemOpen
  \bibfield  {author} {\bibinfo {author} {\bibfnamefont {A.~M.}\ \bibnamefont
  {Essin}}\ and\ \bibinfo {author} {\bibfnamefont {M.}~\bibnamefont
  {Hermele}},\ }\href {\doibase 10.1103/PhysRevB.87.104406} {\bibfield
  {journal} {\bibinfo  {journal} {Phys. Rev. B}\ }\textbf {\bibinfo {volume}
  {87}},\ \bibinfo {pages} {104406} (\bibinfo {year} {2013})}\BibitemShut
  {NoStop}%
\bibitem [{\citenamefont {Barkeshli}\ \emph {et~al.}()\citenamefont
  {Barkeshli}, \citenamefont {Bonderson}, \citenamefont {Cheng},\ and\
  \citenamefont {Wang}}]{BarkeshliX}%
  \BibitemOpen
  \bibfield  {author} {\bibinfo {author} {\bibfnamefont {M.}~\bibnamefont
  {Barkeshli}}, \bibinfo {author} {\bibfnamefont {P.}~\bibnamefont
  {Bonderson}}, \bibinfo {author} {\bibfnamefont {M.}~\bibnamefont {Cheng}}, \
  and\ \bibinfo {author} {\bibfnamefont {Z.}~\bibnamefont {Wang}},\ }\href@noop
  {} {\ }\Eprint {http://arxiv.org/abs/1410.4540} {arXiv:1410.4540
  [cond-mat.str-el]} \BibitemShut {NoStop}%
\bibitem [{\citenamefont {Fidkowski}\ \emph {et~al.}()\citenamefont
  {Fidkowski}, \citenamefont {Lindner},\ and\ \citenamefont
  {Kitaev}}]{Fidkowski_unpub}%
  \BibitemOpen
  \bibfield  {author} {\bibinfo {author} {\bibfnamefont {L.}~\bibnamefont
  {Fidkowski}}, \bibinfo {author} {\bibfnamefont {N.}~\bibnamefont {Lindner}},
  \ and\ \bibinfo {author} {\bibfnamefont {A.}~\bibnamefont {Kitaev}},\
  }\href@noop {} {}\bibinfo {note} {Unpublished}\BibitemShut {NoStop}%
\bibitem [{\citenamefont {Tarantino}\ \emph {et~al.}()\citenamefont
  {Tarantino}, \citenamefont {Lindner},\ and\ \citenamefont
  {Fidkowski}}]{Tarantino_arxiv}%
  \BibitemOpen
  \bibfield  {author} {\bibinfo {author} {\bibfnamefont {N.}~\bibnamefont
  {Tarantino}}, \bibinfo {author} {\bibfnamefont {N.}~\bibnamefont {Lindner}},
  \ and\ \bibinfo {author} {\bibfnamefont {L.}~\bibnamefont {Fidkowski}},\
  }\href@noop {} {}\Eprint {http://arxiv.org/abs/1506.06754} {arXiv:1506.06754
  [cond-mat.str-el]} \BibitemShut {NoStop}%
\bibitem [{\citenamefont {Teo}\ \emph {et~al.}(2015)\citenamefont {Teo},
  \citenamefont {Hughes},\ and\ \citenamefont {Fradkin}}]{Teo2015}%
  \BibitemOpen
  \bibfield  {author} {\bibinfo {author} {\bibfnamefont {J.~C.~Y.}\
  \bibnamefont {Teo}}, \bibinfo {author} {\bibfnamefont {T.~L.}\ \bibnamefont
  {Hughes}}, \ and\ \bibinfo {author} {\bibfnamefont {E.}~\bibnamefont
  {Fradkin}},\ }\href@noop {} {\bibfield  {journal} {\bibinfo  {journal} {Ann.
  Phys.}\ }\textbf {\bibinfo {volume} {360}},\ \bibinfo {pages} {349} (\bibinfo
  {year} {2015})}\BibitemShut {NoStop}%
\bibitem [{\citenamefont {Lu}\ and\ \citenamefont {Vishwanath}()}]{Lu2013}%
  \BibitemOpen
  \bibfield  {author} {\bibinfo {author} {\bibfnamefont {Y.-M.}\ \bibnamefont
  {Lu}}\ and\ \bibinfo {author} {\bibfnamefont {A.}~\bibnamefont
  {Vishwanath}},\ }\href@noop {} {\ }\Eprint {http://arxiv.org/abs/1302.2634}
  {arXiv:1302.2634 [cond-mat.str-el]} \BibitemShut {NoStop}%
\bibitem [{\citenamefont {Hung}\ and\ \citenamefont {Wan}(2013)}]{HungPRB2013}%
  \BibitemOpen
  \bibfield  {author} {\bibinfo {author} {\bibfnamefont {L.-Y.}\ \bibnamefont
  {Hung}}\ and\ \bibinfo {author} {\bibfnamefont {Y.}~\bibnamefont {Wan}},\
  }\href {\doibase 10.1103/PhysRevB.87.195103} {\bibfield  {journal} {\bibinfo
  {journal} {Phys. Rev. B}\ }\textbf {\bibinfo {volume} {87}},\ \bibinfo
  {pages} {195103} (\bibinfo {year} {2013})}\BibitemShut {NoStop}%
\bibitem [{\citenamefont {Laughlin}(1983)}]{LaughlinFQHE}%
  \BibitemOpen
  \bibfield  {author} {\bibinfo {author} {\bibfnamefont {R.~B.}\ \bibnamefont
  {Laughlin}},\ }\href {\doibase 10.1103/PhysRevLett.50.1395} {\bibfield
  {journal} {\bibinfo  {journal} {Phys. Rev. Lett.}\ }\textbf {\bibinfo
  {volume} {50}},\ \bibinfo {pages} {1395} (\bibinfo {year}
  {1983})}\BibitemShut {NoStop}%
\bibitem [{\citenamefont {Kalmeyer}\ and\ \citenamefont
  {Laughlin}(1987)}]{KalmeyerCSL1987}%
  \BibitemOpen
  \bibfield  {author} {\bibinfo {author} {\bibfnamefont {V.}~\bibnamefont
  {Kalmeyer}}\ and\ \bibinfo {author} {\bibfnamefont {R.~B.}\ \bibnamefont
  {Laughlin}},\ }\href {\doibase 10.1103/PhysRevLett.59.2095} {\bibfield
  {journal} {\bibinfo  {journal} {Phys. Rev. Lett.}\ }\textbf {\bibinfo
  {volume} {59}},\ \bibinfo {pages} {2095} (\bibinfo {year}
  {1987})}\BibitemShut {NoStop}%
\bibitem [{\citenamefont {Wen}(2002)}]{wenpsg}%
  \BibitemOpen
  \bibfield  {author} {\bibinfo {author} {\bibfnamefont {X.-G.}\ \bibnamefont
  {Wen}},\ }\href {\doibase 10.1103/PhysRevB.65.165113} {\bibfield  {journal}
  {\bibinfo  {journal} {Phys. Rev. B}\ }\textbf {\bibinfo {volume} {65}},\
  \bibinfo {pages} {165113} (\bibinfo {year} {2002})}\BibitemShut {NoStop}%
\bibitem [{\citenamefont {Yao}\ \emph {et~al.}()\citenamefont {Yao},
  \citenamefont {Fu},\ and\ \citenamefont {Qi}}]{YaoFuQi2010X}%
  \BibitemOpen
  \bibfield  {author} {\bibinfo {author} {\bibfnamefont {H.}~\bibnamefont
  {Yao}}, \bibinfo {author} {\bibfnamefont {L.}~\bibnamefont {Fu}}, \ and\
  \bibinfo {author} {\bibfnamefont {X.-L.}\ \bibnamefont {Qi}},\ }\href@noop {}
  {\ }\Eprint {http://arxiv.org/abs/1012.4470} {arXiv:1012.4470
  [cond-mat.str-el]} \BibitemShut {NoStop}%
\bibitem [{\citenamefont {de~Picciotto}\ \emph {et~al.}(1997)\citenamefont
  {de~Picciotto}, \citenamefont {Reznikov}, \citenamefont {Heiblum},
  \citenamefont {Umansky}, \citenamefont {Bunin},\ and\ \citenamefont
  {Mahalu}}]{Picciotto1997}%
  \BibitemOpen
  \bibfield  {author} {\bibinfo {author} {\bibfnamefont {R.}~\bibnamefont
  {de~Picciotto}}, \bibinfo {author} {\bibfnamefont {M.}~\bibnamefont
  {Reznikov}}, \bibinfo {author} {\bibfnamefont {M.}~\bibnamefont {Heiblum}},
  \bibinfo {author} {\bibfnamefont {V.}~\bibnamefont {Umansky}}, \bibinfo
  {author} {\bibfnamefont {G.}~\bibnamefont {Bunin}}, \ and\ \bibinfo {author}
  {\bibfnamefont {D.}~\bibnamefont {Mahalu}},\ }\href {\doibase 10.1038/38241}
  {\bibfield  {journal} {\bibinfo  {journal} {Nature}\ }\textbf {\bibinfo
  {volume} {389}},\ \bibinfo {pages} {162} (\bibinfo {year}
  {1997})}\BibitemShut {NoStop}%
\bibitem [{\citenamefont {Saminadayar}\ \emph {et~al.}(1997)\citenamefont
  {Saminadayar}, \citenamefont {Glattli}, \citenamefont {Jin},\ and\
  \citenamefont {Etienne}}]{Saminadayar1997}%
  \BibitemOpen
  \bibfield  {author} {\bibinfo {author} {\bibfnamefont {L.}~\bibnamefont
  {Saminadayar}}, \bibinfo {author} {\bibfnamefont {D.~C.}\ \bibnamefont
  {Glattli}}, \bibinfo {author} {\bibfnamefont {Y.}~\bibnamefont {Jin}}, \ and\
  \bibinfo {author} {\bibfnamefont {B.}~\bibnamefont {Etienne}},\ }\href
  {\doibase 10.1103/PhysRevLett.79.2526} {\bibfield  {journal} {\bibinfo
  {journal} {Phys. Rev. Lett.}\ }\textbf {\bibinfo {volume} {79}},\ \bibinfo
  {pages} {2526} (\bibinfo {year} {1997})}\BibitemShut {NoStop}%
\bibitem [{\citenamefont {Han}\ \emph {et~al.}(2012)\citenamefont {Han},
  \citenamefont {Helton}, \citenamefont {Chu}, \citenamefont {Nocera},
  \citenamefont {Rodriguez-Rivera}, \citenamefont {Broholm},\ and\
  \citenamefont {Lee}}]{Han2012}%
  \BibitemOpen
  \bibfield  {author} {\bibinfo {author} {\bibfnamefont {T.-H.}\ \bibnamefont
  {Han}}, \bibinfo {author} {\bibfnamefont {J.~S.}\ \bibnamefont {Helton}},
  \bibinfo {author} {\bibfnamefont {S.}~\bibnamefont {Chu}}, \bibinfo {author}
  {\bibfnamefont {D.~G.}\ \bibnamefont {Nocera}}, \bibinfo {author}
  {\bibfnamefont {J.~A.}\ \bibnamefont {Rodriguez-Rivera}}, \bibinfo {author}
  {\bibfnamefont {C.}~\bibnamefont {Broholm}}, \ and\ \bibinfo {author}
  {\bibfnamefont {Y.~S.}\ \bibnamefont {Lee}},\ }\href {\doibase
  10.1038/nature11659} {\bibfield  {journal} {\bibinfo  {journal} {Nature}\
  }\textbf {\bibinfo {volume} {492}},\ \bibinfo {pages} {406} (\bibinfo {year}
  {2012})}\BibitemShut {NoStop}%
\bibitem [{\citenamefont {Wang}\ and\ \citenamefont
  {Vishwanath}(2006)}]{wang2007}%
  \BibitemOpen
  \bibfield  {author} {\bibinfo {author} {\bibfnamefont {F.}~\bibnamefont
  {Wang}}\ and\ \bibinfo {author} {\bibfnamefont {A.}~\bibnamefont
  {Vishwanath}},\ }\href {\doibase 10.1103/PhysRevB.74.174423} {\bibfield
  {journal} {\bibinfo  {journal} {Phys. Rev. B}\ }\textbf {\bibinfo {volume}
  {74}},\ \bibinfo {pages} {174423} (\bibinfo {year} {2006})}\BibitemShut
  {NoStop}%
\bibitem [{\citenamefont {Jiang}\ and\ \citenamefont {Ran}()}]{SJiangTPS2015X}%
  \BibitemOpen
  \bibfield  {author} {\bibinfo {author} {\bibfnamefont {S.}~\bibnamefont
  {Jiang}}\ and\ \bibinfo {author} {\bibfnamefont {Y.}~\bibnamefont {Ran}},\
  }\href@noop {} {\ }\Eprint {http://arxiv.org/abs/1505.03171}
  {arXiv:1505.03171 [cond-mat.str-el]} \BibitemShut {NoStop}%
\bibitem [{\citenamefont {Lu}\ \emph {et~al.}(2011)\citenamefont {Lu},
  \citenamefont {Ran},\ and\ \citenamefont {Lee}}]{YMLuKagomePSG2011}%
  \BibitemOpen
  \bibfield  {author} {\bibinfo {author} {\bibfnamefont {Y.-M.}\ \bibnamefont
  {Lu}}, \bibinfo {author} {\bibfnamefont {Y.}~\bibnamefont {Ran}}, \ and\
  \bibinfo {author} {\bibfnamefont {P.~A.}\ \bibnamefont {Lee}},\ }\href
  {\doibase 10.1103/PhysRevB.83.224413} {\bibfield  {journal} {\bibinfo
  {journal} {Phys. Rev. B}\ }\textbf {\bibinfo {volume} {83}},\ \bibinfo
  {pages} {224413} (\bibinfo {year} {2011})}\BibitemShut {NoStop}%
\bibitem [{\citenamefont {Lu}\ \emph {et~al.}()\citenamefont {Lu},
  \citenamefont {Cho},\ and\ \citenamefont {Vishwanath}}]{LuBFU}%
  \BibitemOpen
  \bibfield  {author} {\bibinfo {author} {\bibfnamefont {Y.-M.}\ \bibnamefont
  {Lu}}, \bibinfo {author} {\bibfnamefont {G.~Y.}\ \bibnamefont {Cho}}, \ and\
  \bibinfo {author} {\bibfnamefont {A.}~\bibnamefont {Vishwanath}},\
  }\href@noop {} {\ }\Eprint {http://arxiv.org/abs/1403.0575} {arXiv:1403.0575
  [cond-mat.str-el]} \BibitemShut {NoStop}%
\bibitem [{\citenamefont {Zheng}\ \emph {et~al.}()\citenamefont {Zheng},
  \citenamefont {Mei},\ and\ \citenamefont {Qi}}]{Zheng_arxiv}%
  \BibitemOpen
  \bibfield  {author} {\bibinfo {author} {\bibfnamefont {W.}~\bibnamefont
  {Zheng}}, \bibinfo {author} {\bibfnamefont {J.-W.}\ \bibnamefont {Mei}}, \
  and\ \bibinfo {author} {\bibfnamefont {Y.}~\bibnamefont {Qi}},\ }\href@noop
  {} {\ }\Eprint {http://arxiv.org/abs/1505.05351} {arXiv:1505.05351
  [cond-mat.str-el]} \BibitemShut {NoStop}%
\bibitem [{\citenamefont {Lu}()}]{YMLuTri2015X}%
  \BibitemOpen
  \bibfield  {author} {\bibinfo {author} {\bibfnamefont {Y.-M.}\ \bibnamefont
  {Lu}},\ }\href@noop {} {\ }\Eprint {http://arxiv.org/abs/1505.06495}
  {arXiv:1505.06495 [cond-mat.str-el]} \BibitemShut {NoStop}%
\bibitem [{\citenamefont {Yang}\ and\ \citenamefont {Wang}()}]{FWangSq2015X}%
  \BibitemOpen
  \bibfield  {author} {\bibinfo {author} {\bibfnamefont {X.}~\bibnamefont
  {Yang}}\ and\ \bibinfo {author} {\bibfnamefont {F.}~\bibnamefont {Wang}},\
  }\href@noop {} {\ }\Eprint {http://arxiv.org/abs/1507.07621}
  {arXiv:1507.07621 [cond-mat.str-el]} \BibitemShut {NoStop}%
\bibitem [{\citenamefont {Vishwanath}\ and\ \citenamefont
  {Senthil}(2013)}]{VishwanathPRX2013}%
  \BibitemOpen
  \bibfield  {author} {\bibinfo {author} {\bibfnamefont {A.}~\bibnamefont
  {Vishwanath}}\ and\ \bibinfo {author} {\bibfnamefont {T.}~\bibnamefont
  {Senthil}},\ }\href {\doibase 10.1103/PhysRevX.3.011016} {\bibfield
  {journal} {\bibinfo  {journal} {Phys. Rev. X}\ }\textbf {\bibinfo {volume}
  {3}},\ \bibinfo {pages} {011016} (\bibinfo {year} {2013})}\BibitemShut
  {NoStop}%
\bibitem [{\citenamefont {Wang}\ and\ \citenamefont
  {Senthil}(2013)}]{CWangETMT2013}%
  \BibitemOpen
  \bibfield  {author} {\bibinfo {author} {\bibfnamefont {C.}~\bibnamefont
  {Wang}}\ and\ \bibinfo {author} {\bibfnamefont {T.}~\bibnamefont {Senthil}},\
  }\href {\doibase 10.1103/PhysRevB.87.235122} {\bibfield  {journal} {\bibinfo
  {journal} {Phys. Rev. B}\ }\textbf {\bibinfo {volume} {87}},\ \bibinfo
  {pages} {235122} (\bibinfo {year} {2013})}\BibitemShut {NoStop}%
\bibitem [{\citenamefont {Qi}\ and\ \citenamefont {Fu}()}]{Qi_unpub}%
  \BibitemOpen
  \bibfield  {author} {\bibinfo {author} {\bibfnamefont {Y.}~\bibnamefont
  {Qi}}\ and\ \bibinfo {author} {\bibfnamefont {L.}~\bibnamefont {Fu}},\
  }\href@noop {} {}\Eprint {http://arxiv.org/abs/1505.06201} {arXiv:1505.06201
  [cond-mat.str-el]} \BibitemShut {NoStop}%
\bibitem [{\citenamefont {Hermele}\ and\ \citenamefont {Chen}()}]{HermeleFFAT}%
  \BibitemOpen
  \bibfield  {author} {\bibinfo {author} {\bibfnamefont {M.}~\bibnamefont
  {Hermele}}\ and\ \bibinfo {author} {\bibfnamefont {X.}~\bibnamefont {Chen}},\
  }\href@noop {} {\ }\Eprint {http://arxiv.org/abs/1508.00573}
  {arXiv:1508.00573 [cond-mat.str-el]} \BibitemShut {NoStop}%
\bibitem [{\citenamefont {Chen}\ \emph {et~al.}(2013)\citenamefont {Chen},
  \citenamefont {Gu}, \citenamefont {Liu},\ and\ \citenamefont
  {Wen}}]{XChenSPT}%
  \BibitemOpen
  \bibfield  {author} {\bibinfo {author} {\bibfnamefont {X.}~\bibnamefont
  {Chen}}, \bibinfo {author} {\bibfnamefont {Z.-C.}\ \bibnamefont {Gu}},
  \bibinfo {author} {\bibfnamefont {Z.-X.}\ \bibnamefont {Liu}}, \ and\
  \bibinfo {author} {\bibfnamefont {X.-G.}\ \bibnamefont {Wen}},\ }\href
  {\doibase 10.1103/PhysRevB.87.155114} {\bibfield  {journal} {\bibinfo
  {journal} {Phys. Rev. B}\ }\textbf {\bibinfo {volume} {87}},\ \bibinfo
  {pages} {155114} (\bibinfo {year} {2013})}\BibitemShut {NoStop}%
\bibitem [{\citenamefont {Schnyder}\ \emph {et~al.}(2008)\citenamefont
  {Schnyder}, \citenamefont {Ryu}, \citenamefont {Furusaki},\ and\
  \citenamefont {Ludwig}}]{SchnyderTSC2008}%
  \BibitemOpen
  \bibfield  {author} {\bibinfo {author} {\bibfnamefont {A.~P.}\ \bibnamefont
  {Schnyder}}, \bibinfo {author} {\bibfnamefont {S.}~\bibnamefont {Ryu}},
  \bibinfo {author} {\bibfnamefont {A.}~\bibnamefont {Furusaki}}, \ and\
  \bibinfo {author} {\bibfnamefont {A.~W.~W.}\ \bibnamefont {Ludwig}},\ }\href
  {\doibase 10.1103/PhysRevB.78.195125} {\bibfield  {journal} {\bibinfo
  {journal} {Phys. Rev. B}\ }\textbf {\bibinfo {volume} {78}},\ \bibinfo
  {pages} {195125} (\bibinfo {year} {2008})}\BibitemShut {NoStop}%
\bibitem [{\citenamefont {Fu}(2011)}]{FuTCI2011}%
  \BibitemOpen
  \bibfield  {author} {\bibinfo {author} {\bibfnamefont {L.}~\bibnamefont
  {Fu}},\ }\href {\doibase 10.1103/PhysRevLett.106.106802} {\bibfield
  {journal} {\bibinfo  {journal} {Phys. Rev. Lett.}\ }\textbf {\bibinfo
  {volume} {106}},\ \bibinfo {pages} {106802} (\bibinfo {year}
  {2011})}\BibitemShut {NoStop}%
\bibitem [{\citenamefont {Chen}\ \emph {et~al.}()\citenamefont {Chen},
  \citenamefont {Burnell}, \citenamefont {Vishwanath},\ and\ \citenamefont
  {Fidkowski}}]{Chen2014}%
  \BibitemOpen
  \bibfield  {author} {\bibinfo {author} {\bibfnamefont {X.}~\bibnamefont
  {Chen}}, \bibinfo {author} {\bibfnamefont {F.~J.}\ \bibnamefont {Burnell}},
  \bibinfo {author} {\bibfnamefont {A.}~\bibnamefont {Vishwanath}}, \ and\
  \bibinfo {author} {\bibfnamefont {L.}~\bibnamefont {Fidkowski}},\ }\href@noop
  {} {}\Eprint {http://arxiv.org/abs/1403.6491} {arXiv:1403.6491
  [cond-mat.str-el]} \BibitemShut {NoStop}%
\bibitem [{\citenamefont {Yan}\ \emph {et~al.}()\citenamefont {Yan},
  \citenamefont {Huse},\ and\ \citenamefont {White}}]{YanScience}%
  \BibitemOpen
  \bibfield  {author} {\bibinfo {author} {\bibfnamefont {S.}~\bibnamefont
  {Yan}}, \bibinfo {author} {\bibfnamefont {D.~A.}\ \bibnamefont {Huse}}, \
  and\ \bibinfo {author} {\bibfnamefont {S.~R.}\ \bibnamefont {White}},\
  }\href@noop {} {\bibfield  {journal} {\bibinfo  {journal} {Science}\ }\textbf
  {\bibinfo {volume} {332}},\ \bibinfo {pages} {1173}}\BibitemShut {NoStop}%
\bibitem [{\citenamefont {Jiang}\ \emph {et~al.}()\citenamefont {Jiang},
  \citenamefont {Wang},\ and\ \citenamefont {Balents}}]{JiangNatPhys}%
  \BibitemOpen
  \bibfield  {author} {\bibinfo {author} {\bibfnamefont {H.-C.}\ \bibnamefont
  {Jiang}}, \bibinfo {author} {\bibfnamefont {Z.}~\bibnamefont {Wang}}, \ and\
  \bibinfo {author} {\bibfnamefont {L.}~\bibnamefont {Balents}},\ }\href@noop
  {} {\bibfield  {journal} {\bibinfo  {journal} {Nature Phys.}\ }\textbf
  {\bibinfo {volume} {8}},\ \bibinfo {pages} {902}}\BibitemShut {NoStop}%
\bibitem [{\citenamefont {Depenbrock}\ \emph {et~al.}(2012)\citenamefont
  {Depenbrock}, \citenamefont {McCulloch},\ and\ \citenamefont
  {Schollw\"ock}}]{DepenbrockPRL2012}%
  \BibitemOpen
  \bibfield  {author} {\bibinfo {author} {\bibfnamefont {S.}~\bibnamefont
  {Depenbrock}}, \bibinfo {author} {\bibfnamefont {I.~P.}\ \bibnamefont
  {McCulloch}}, \ and\ \bibinfo {author} {\bibfnamefont {U.}~\bibnamefont
  {Schollw\"ock}},\ }\href {\doibase 10.1103/PhysRevLett.109.067201} {\bibfield
   {journal} {\bibinfo  {journal} {Phys. Rev. Lett.}\ }\textbf {\bibinfo
  {volume} {109}},\ \bibinfo {pages} {067201} (\bibinfo {year}
  {2012})}\BibitemShut {NoStop}%
\bibitem [{\citenamefont {Iqbal}\ \emph {et~al.}(2011)\citenamefont {Iqbal},
  \citenamefont {Becca},\ and\ \citenamefont {Poilblanc}}]{IqbalKagome2011}%
  \BibitemOpen
  \bibfield  {author} {\bibinfo {author} {\bibfnamefont {Y.}~\bibnamefont
  {Iqbal}}, \bibinfo {author} {\bibfnamefont {F.}~\bibnamefont {Becca}}, \ and\
  \bibinfo {author} {\bibfnamefont {D.}~\bibnamefont {Poilblanc}},\ }\href
  {\doibase 10.1103/PhysRevB.84.020407} {\bibfield  {journal} {\bibinfo
  {journal} {Phys. Rev. B}\ }\textbf {\bibinfo {volume} {84}},\ \bibinfo
  {pages} {020407(R)} (\bibinfo {year} {2011})}\BibitemShut {NoStop}%
\bibitem [{\citenamefont {Fu}\ \emph {et~al.}()\citenamefont {Fu},
  \citenamefont {Imai}, \citenamefont {Han},\ and\ \citenamefont
  {Lee}}]{YSLeeNMR_aps}%
  \BibitemOpen
  \bibfield  {author} {\bibinfo {author} {\bibfnamefont {M.}~\bibnamefont
  {Fu}}, \bibinfo {author} {\bibfnamefont {T.}~\bibnamefont {Imai}}, \bibinfo
  {author} {\bibfnamefont {T.}~\bibnamefont {Han}}, \ and\ \bibinfo {author}
  {\bibfnamefont {Y.~S.}\ \bibnamefont {Lee}},\ }\href
  {http://meetings.aps.org/link/BAPS.2015.MAR.Z28.11} {\enquote {\bibinfo
  {title} {$^{17}${O} single crystal {NMR} evidence for a gapped spin-liquid
  ground state in the {$S=\frac12$} kagome lattice {ZnCu$_3$(OH)$_6$Cl$_2$}},}\
  }\bibinfo {note} {APS March Meeting 2015.}\BibitemShut {Stop}%
\bibitem [{\citenamefont {Balents}\ \emph {et~al.}(2002)\citenamefont
  {Balents}, \citenamefont {Fisher},\ and\ \citenamefont
  {Girvin}}]{BFGZ2SL2002}%
  \BibitemOpen
  \bibfield  {author} {\bibinfo {author} {\bibfnamefont {L.}~\bibnamefont
  {Balents}}, \bibinfo {author} {\bibfnamefont {M.~P.~A.}\ \bibnamefont
  {Fisher}}, \ and\ \bibinfo {author} {\bibfnamefont {S.~M.}\ \bibnamefont
  {Girvin}},\ }\href {\doibase 10.1103/PhysRevB.65.224412} {\bibfield
  {journal} {\bibinfo  {journal} {Phys. Rev. B}\ }\textbf {\bibinfo {volume}
  {65}},\ \bibinfo {pages} {224412} (\bibinfo {year} {2002})}\BibitemShut
  {NoStop}%
\bibitem [{\citenamefont {Metlitski}\ \emph {et~al.}()\citenamefont
  {Metlitski}, \citenamefont {Kane},\ and\ \citenamefont
  {Fisher}}]{metlitski2013}%
  \BibitemOpen
  \bibfield  {author} {\bibinfo {author} {\bibfnamefont {M.~A.}\ \bibnamefont
  {Metlitski}}, \bibinfo {author} {\bibfnamefont {C.~L.}\ \bibnamefont {Kane}},
  \ and\ \bibinfo {author} {\bibfnamefont {M.~P.~A.}\ \bibnamefont {Fisher}},\
  }\href@noop {} {}\Eprint {http://arxiv.org/abs/1306.3286} {arXiv:1306.3286
  [cond-mat.str-el]} \BibitemShut {NoStop}%
\bibitem [{\citenamefont {Zhu}\ and\ \citenamefont {White}(2015)}]{ZhuPRB2015}%
  \BibitemOpen
  \bibfield  {author} {\bibinfo {author} {\bibfnamefont {Z.}~\bibnamefont
  {Zhu}}\ and\ \bibinfo {author} {\bibfnamefont {S.~R.}\ \bibnamefont
  {White}},\ }\href {\doibase 10.1103/PhysRevB.92.041105} {\bibfield  {journal}
  {\bibinfo  {journal} {Phys. Rev. B}\ }\textbf {\bibinfo {volume} {92}},\
  \bibinfo {pages} {041105(R)} (\bibinfo {year} {2015})}\BibitemShut {NoStop}%
\bibitem [{\citenamefont {Hu}\ \emph {et~al.}()\citenamefont {Hu},
  \citenamefont {Gong}, \citenamefont {Zhu},\ and\ \citenamefont
  {Sheng}}]{WJHuTriZ2SL2015}%
  \BibitemOpen
  \bibfield  {author} {\bibinfo {author} {\bibfnamefont {W.-J.}\ \bibnamefont
  {Hu}}, \bibinfo {author} {\bibfnamefont {S.-S.}\ \bibnamefont {Gong}},
  \bibinfo {author} {\bibfnamefont {W.}~\bibnamefont {Zhu}}, \ and\ \bibinfo
  {author} {\bibfnamefont {D.~N.}\ \bibnamefont {Sheng}},\ }\href@noop {} {\
  }\Eprint {http://arxiv.org/abs/1504.00654} {arXiv:1504.00654
  [cond-mat.str-el]} \BibitemShut {NoStop}%
\bibitem [{\citenamefont {Tay}\ and\ \citenamefont
  {Motrunich}(2011)}]{TayVMC2011}%
  \BibitemOpen
  \bibfield  {author} {\bibinfo {author} {\bibfnamefont {T.}~\bibnamefont
  {Tay}}\ and\ \bibinfo {author} {\bibfnamefont {O.~I.}\ \bibnamefont
  {Motrunich}},\ }\href {\doibase 10.1103/PhysRevB.84.020404} {\bibfield
  {journal} {\bibinfo  {journal} {Phys. Rev. B}\ }\textbf {\bibinfo {volume}
  {84}},\ \bibinfo {pages} {020404(R)} (\bibinfo {year} {2011})}\BibitemShut
  {NoStop}%
\bibitem [{\citenamefont {Rousochatzakis}\ \emph {et~al.}(2014)\citenamefont
  {Rousochatzakis}, \citenamefont {Wan}, \citenamefont {Tchernyshyov},\ and\
  \citenamefont {Mila}}]{Rousochatzakis2014}%
  \BibitemOpen
  \bibfield  {author} {\bibinfo {author} {\bibfnamefont {I.}~\bibnamefont
  {Rousochatzakis}}, \bibinfo {author} {\bibfnamefont {Y.}~\bibnamefont {Wan}},
  \bibinfo {author} {\bibfnamefont {O.}~\bibnamefont {Tchernyshyov}}, \ and\
  \bibinfo {author} {\bibfnamefont {F.}~\bibnamefont {Mila}},\ }\href {\doibase
  10.1103/PhysRevB.90.100406} {\bibfield  {journal} {\bibinfo  {journal} {Phys.
  Rev. B}\ }\textbf {\bibinfo {volume} {90}},\ \bibinfo {pages} {100406(R)}
  (\bibinfo {year} {2014})}\BibitemShut {NoStop}%
\bibitem [{\citenamefont {Lieb}\ \emph {et~al.}(1961)\citenamefont {Lieb},
  \citenamefont {Schultz},\ and\ \citenamefont {Mattis}}]{LSM}%
  \BibitemOpen
  \bibfield  {author} {\bibinfo {author} {\bibfnamefont {E.}~\bibnamefont
  {Lieb}}, \bibinfo {author} {\bibfnamefont {T.}~\bibnamefont {Schultz}}, \
  and\ \bibinfo {author} {\bibfnamefont {D.}~\bibnamefont {Mattis}},\
  }\href@noop {} {\bibfield  {journal} {\bibinfo  {journal} {Ann. Phys.}\
  }\textbf {\bibinfo {volume} {16}},\ \bibinfo {pages} {407} (\bibinfo {year}
  {1961})}\BibitemShut {NoStop}%
\bibitem [{Note1()}]{Note1}%
  \BibitemOpen
  \bibinfo {note} {We notice that the adiabatic insertion is not essential for
  the flux-fusion argument to work, and it can be generalized to a discrete
  subgroup of the $\protect \mathrm {U}(1)$ symmetry, see Ref. [\protect
  \rev@citealp {HermeleFFAT}].}\BibitemShut {Stop}%
\bibitem [{\citenamefont {Qi}\ and\ \citenamefont {Fu}(2015)}]{QiCSF}%
  \BibitemOpen
  \bibfield  {author} {\bibinfo {author} {\bibfnamefont {Y.}~\bibnamefont
  {Qi}}\ and\ \bibinfo {author} {\bibfnamefont {L.}~\bibnamefont {Fu}},\ }\href
  {\doibase 10.1103/PhysRevB.91.100401} {\bibfield  {journal} {\bibinfo
  {journal} {Phys. Rev. B}\ }\textbf {\bibinfo {volume} {91}},\ \bibinfo
  {pages} {100401(R)} (\bibinfo {year} {2015})}\BibitemShut {NoStop}%
\bibitem [{\citenamefont {Zaletel}\ \emph {et~al.}()\citenamefont {Zaletel},
  \citenamefont {Lu},\ and\ \citenamefont {Vishwanath}}]{ZLVPSG}%
  \BibitemOpen
  \bibfield  {author} {\bibinfo {author} {\bibfnamefont {M.}~\bibnamefont
  {Zaletel}}, \bibinfo {author} {\bibfnamefont {Y.-M.}\ \bibnamefont {Lu}}, \
  and\ \bibinfo {author} {\bibfnamefont {A.}~\bibnamefont {Vishwanath}},\
  }\href@noop {} {\ }\Eprint {http://arxiv.org/abs/1501.01395}
  {arXiv:1501.01395 [cond-mat.str-el]} \BibitemShut {NoStop}%
\bibitem [{\citenamefont {Sachdev}\ and\ \citenamefont
  {Vojta}(2000)}]{SachdevFFIM1999}%
  \BibitemOpen
  \bibfield  {author} {\bibinfo {author} {\bibfnamefont {S.}~\bibnamefont
  {Sachdev}}\ and\ \bibinfo {author} {\bibfnamefont {M.}~\bibnamefont
  {Vojta}},\ }\href@noop {} {\bibfield  {journal} {\bibinfo  {journal} {J.
  Phys. Soc. Japan}\ }\textbf {\bibinfo {volume} {69 Suppl B}},\ \bibinfo
  {pages} {1} (\bibinfo {year} {2000})},\ \Eprint
  {http://arxiv.org/abs/cond-mat/9910231} {arXiv:cond-mat/9910231
  [cond-mat.str-el]} \BibitemShut {NoStop}%
\bibitem [{\citenamefont {Senthil}\ and\ \citenamefont
  {Fisher}(2000)}]{SenthilZ2SL2000}%
  \BibitemOpen
  \bibfield  {author} {\bibinfo {author} {\bibfnamefont {T.}~\bibnamefont
  {Senthil}}\ and\ \bibinfo {author} {\bibfnamefont {M.~P.~A.}\ \bibnamefont
  {Fisher}},\ }\href {\doibase 10.1103/PhysRevB.62.7850} {\bibfield  {journal}
  {\bibinfo  {journal} {Phys. Rev. B}\ }\textbf {\bibinfo {volume} {62}},\
  \bibinfo {pages} {7850} (\bibinfo {year} {2000})}\BibitemShut {NoStop}%
\bibitem [{\citenamefont {Moessner}\ and\ \citenamefont
  {Sondhi}(2001)}]{MoessnerTri2001}%
  \BibitemOpen
  \bibfield  {author} {\bibinfo {author} {\bibfnamefont {R.}~\bibnamefont
  {Moessner}}\ and\ \bibinfo {author} {\bibfnamefont {S.~L.}\ \bibnamefont
  {Sondhi}},\ }\href {\doibase 10.1103/PhysRevLett.86.1881} {\bibfield
  {journal} {\bibinfo  {journal} {Phys. Rev. Lett.}\ }\textbf {\bibinfo
  {volume} {86}},\ \bibinfo {pages} {1881} (\bibinfo {year}
  {2001})}\BibitemShut {NoStop}%
\bibitem [{\citenamefont {Moessner}\ \emph {et~al.}(2001)\citenamefont
  {Moessner}, \citenamefont {Sondhi},\ and\ \citenamefont
  {Fradkin}}]{MoessnerZ2Gauge2001}%
  \BibitemOpen
  \bibfield  {author} {\bibinfo {author} {\bibfnamefont {R.}~\bibnamefont
  {Moessner}}, \bibinfo {author} {\bibfnamefont {S.~L.}\ \bibnamefont
  {Sondhi}}, \ and\ \bibinfo {author} {\bibfnamefont {E.}~\bibnamefont
  {Fradkin}},\ }\href {\doibase 10.1103/PhysRevB.65.024504} {\bibfield
  {journal} {\bibinfo  {journal} {Phys. Rev. B}\ }\textbf {\bibinfo {volume}
  {65}},\ \bibinfo {pages} {024504} (\bibinfo {year} {2001})}\BibitemShut
  {NoStop}%
\bibitem [{Note2()}]{Note2}%
  \BibitemOpen
  \bibinfo {note} {To better compare the two phases we can perturb the
  antiferomagnetic Heisenberg model slightly by introducing easy-axis
  anisotropy. Such a nearest-neighbor XXZ model was studied numerically in Ref.
  \protect \rev@citealp {HePRL2015}, and it was established that the $\protect
  \mathbb {Z}_2$ spin liquid phase persists when the easy-axis anisotropy is
  turned on. Ref. \protect \rev@citealp {HePRL2015} also made the observation
  that the BFG model and the Heisenberg model describe different $\protect
  \mathbb {Z}_2$ spin liquids, but the claimed reason that the spinons in the
  BFG model are Kramers singlet is incorrect.}\BibitemShut {Stop}%
\bibitem [{\citenamefont {Sachdev}(1992)}]{sstri}%
  \BibitemOpen
  \bibfield  {author} {\bibinfo {author} {\bibfnamefont {S.}~\bibnamefont
  {Sachdev}},\ }\href {\doibase 10.1103/PhysRevB.45.12377} {\bibfield
  {journal} {\bibinfo  {journal} {Phys. Rev. B}\ }\textbf {\bibinfo {volume}
  {45}},\ \bibinfo {pages} {12377} (\bibinfo {year} {1992})}\BibitemShut
  {NoStop}%
\bibitem [{\citenamefont {He}\ and\ \citenamefont {Chen}(2015)}]{HePRL2015}%
  \BibitemOpen
  \bibfield  {author} {\bibinfo {author} {\bibfnamefont {Y.-C.}\ \bibnamefont
  {He}}\ and\ \bibinfo {author} {\bibfnamefont {Y.}~\bibnamefont {Chen}},\
  }\href {\doibase 10.1103/PhysRevLett.114.037201} {\bibfield  {journal}
  {\bibinfo  {journal} {Phys. Rev. Lett.}\ }\textbf {\bibinfo {volume} {114}},\
  \bibinfo {pages} {037201} (\bibinfo {year} {2015})}\BibitemShut {NoStop}%
\end{thebibliography}%

\clearpage


\section*{Supplementary material}

# Supplemental Material

In this supplemental material we provide technical details of the discussions in the main text. In Sec. I we explain the mathematical details of fractionalization of spin rotational symmetry and crystal symmetries. In Sec. II we discuss the application of flux-fusion anomaly test to site-centered inversions. In Sec. III we derive the details of symmetry fractionalization in the Balents-Fisher-Girvin (BFG) model[10]. In Sec. IV we discuss vison's symmetry fractionalization in gapped $\mathbb{Z}_2$ spin liquids on the square lattice.

## I. FRACTIONALIZATION OF SPIN ROTATIONAL SYMMETRY AND CRYSTAL SYMMETRIES: SO(3) V.S. O(2)

In this section we provide more details on symmetry fractionalization of the symmetry groups of spin liquids, which are a direct product of the spin rotational symmetry, the time-reversal symmetry and crystal symmetries. Particularly we discuss the possibility of fractionalizing the commutation relation between spin rotational symmetry and crystal symmetries, and the projective representation of spin rotational symmetry and crystal symmetries if the former is reduced from the full SO(3) symmetry.

First, we consider the simple case where the spin liquid has a full SO(3) spin rotational symmetry. In this case it is well known that the projective representation of the spin rotational symmetry is classified by $H^2(\mathrm{SO}(3), \mathbb{Z}_2) = \mathbb{Z}_2$, which consists the trivial integer spin representations and the nontrivial half-integer representations. If we enlarge the symmetry group to $\mathrm{SO}(3) \times \mathbb{Z}_2$, where $\mathbb{Z}_2 = \{1, X\}$ represents the time-reversal symmetry or a crystal symmetry, its fractionalization classes are given by the Kunneth formula

$$\begin{aligned} H^2(G \times \mathbb{Z}_2, \mathbb{Z}_2) &= H^2(G, \mathbb{Z}_2) \times \mathbb{Z}_2 \times H^1(G, H^1(\mathbb{Z}_2, \mathbb{Z}_2)) \\ &= H^2(G, \mathbb{Z}_2) \times \mathbb{Z}_2 \times H^1(G, \mathbb{Z}_2), \end{aligned} \quad (1)$$

where $G = \mathrm{SO}(3)$ and we have used $H^1(\mathbb{Z}_2, \mathbb{Z}_2) = \mathbb{Z}_2$ and $H^2(\mathbb{Z}_2, \mathbb{Z}_2) = \mathbb{Z}_2$. In other words, the fractionalization of the symmetry group $\mathrm{SO}(3) \times \mathbb{Z}_2$ are labeled by the combination of the fractionalization of $G$, the fractionalization of of $\mathbb{Z}_2$: $H^2(\mathbb{Z}_2, \mathbb{Z}_2) = \mathbb{Z}_2$, and an extra factor $H^1(G, \mathbb{Z}_2)$, which classifies group homomorphisms from $G$ to $\mathbb{Z}_2$. For a continuous, connected group $G$, the image of any $g \in G$ under the homomorphism can be continuously

connected back to the image of the identity. Since now the images are valued in a discrete group $\mathbb{Z}_2$, we immediately conclude that the homomorphism can only be trivial (since the image of the identity must be identity). More intuitively, we observe that last factor can be thought as the fractionalization of commutation relation between an element $g$ of $G$ and the nontrivial element $X$ of $\mathbb{Z}_2$, $f(g) = gXg^{-1}X^{-1}$, and $f(g)$ takes value in $\pm 1$. Since the values of $f(g)$ are quantized, we must have $f(g_1) = f(g_2)$ if two elements $g_1$ and $g_2$ can be smoothly connected in $G$. Particularly if $g \in G$ is connected to identity we conclude $f(g) = f(1) = +1$. Since $G = \mathrm{SO}(3)$ is connected, all its elements must commute with $X$ in any fractionalization class. We notice that the same argument was given in Ref. [9]. The conclusion is used in the main text to show that $(Xe^{i\pi S^y})^2 = +1$ and $(e^{i\pi S^y})^2 = +1$ implies $X^2 = +1$.

The symmetry fractionalization become more complex if the spin rotational symmetry is reduced to $\mathrm{U}(1) \rtimes \mathbb{Z}_2 = \mathrm{O}(2)$. First, the fractionalization of the spin rotation itself become more complex as $H^2(\mathrm{O}(2), \mathbb{Z}_2) = \mathbb{Z}_2^2$, and are labeled by two independent $\mathbb{Z}_2$ variables denoting $(e^{i\pi S^z})^2 = \pm 1$ and $(e^{i\pi S^y})^2 = \pm 1$, respectively. So if we still assume that the $e$ particle carries a half charge of the $\mathrm{U}(1)$ symmetry $(e^{i\pi S^z})^2 = -1$, the vison can take $(e^{i\pi S^y})^2 = -1$ regardless of whether the $e$ particle takes $(e^{i\pi S^y})^2 = \pm 1$. But with time-reversal symmetry, we can exclude the possibility that both $e$ and $m$ have $(e^{i\pi S^y})^2 = -1$. Second, the spin rotation $e^{i\pi S^y}$ can anticommute with a crystal symmetry $X$ because it is no longer connected to the identity in $\mathrm{O}(2)$. Therefore if the spin rotational symmetry is reduced to $\mathrm{O}(2)$, our argument showing $X^2 = +1$ from $(Xe^{i\pi S^y})^2 = +1$ no longer holds and the vison can take nontrivial symmetry fractionalization.

## II. SITE-CENTERED INVERSION.

In this section we explain that our argument of vison carrying trivial inversion symmetry fractionalization $\omega_I = +1$ only applies to a plaquette-centered inversion $I_p$, not to a site-centered inversion $I_s$, for lattice models with a half-integer spin on each site [1]. The reason that our argument fails for site-centered inversion $I_s$ is that the branch-cut connecting the two fluxes shown in Fig. 1(b) of the main text cannot be chosen to be symmetric under $I_s$. By definition the branch-cut separates sites where the unitary transformation creating the fluxes acts discontinuously and thus cannot go through a site, however without going

though the site at the inversion center it cannot be inversion-symmetric.

To fix this issue, we can choose a branch-cut that is not invariant under inversion, and after the inversion symmetry operation we apply an unitary transformation to restore the original position of the branch-cut. Particularly if we choose a branch-cut such that it and its inversion image encloses only the site of the inversion center, we can apply the unitary transformation $e^{i\phi S^z}$ at the inversion center to restore the branch-cut after the inversion. Therefore the anomaly test works for the combination of inversion symmetry $I_s$ (plus $e^{i\pi S^y}$) and this unitary transformation. If the site at the inversion center carries an integer spin, this unitary transformation becomes trivial at $\phi = 2\pi$ and we still get the result that $I_s^2 = +1$. However, if the site at the inversion center carries a half-integer spin, at $\phi = 2\pi$ this unitary transformation gives an extra $-1$ Berry phase. This explains that in such models vison always carries opposite fractional quantum numbers $I_s^2 = -I_p^2$ [2], and our argument fixes them at $I_p^2 = +1$ and $I_s^2 = -1$, respectively. The relation $I_s^2 = -I_p^2$ can also be derived directly from odd Ising gauge theories and vison's PSG on different lattices [3–8].

Since $I^2$ represents a $2\pi$ rotation, $I_{p/s}^2 = -1$ for the vison can be interpreted as the vison braiding around a background spinon ($e$ or $\psi$) sitting at the inversion center, and $I_p^2 = -I_s^2$ is fixed by the spin-1/2 per unit cell. The anomaly test shows that for SO(3) symmetry, background spinons must sit at the sites. When the spin symmetry is reduced, the position of the background spinons can shift away from the sites to the plaquette centers, as we see in the example of the BFG model discussed in the main text.

## III. SYMMETRY FRACTIONALIZATION IN THE BFG MODEL

In this section we systematically calculate the symmetry fractionalization of the vison in the BFG model [10]. The results obtained here are used in the main text to argue that in this $\mathbb{Z}_2$ spin liquid state the vison carries nontrivial symmetry fractionalization class with $I_p^2 = -1$ and $(\mu T)^2 = -1$. The symmetry group of the BFG model is $O(2) \times \mathbb{Z}_2^T \times p6m$, where the spin rotational symmetry $O(2) = U(1) \rtimes \mathbb{Z}_2$ is generated by $S_z$ and $e^{i\pi S^y}$, respectively. As shown in Table I of the main text, the fractionalization of translation symmetries are labeled by $\omega_{12}$ and it is separated from all other symmetries, so we only need to consider the point group symmetry generated by $\mu$ and $\sigma$.

We start with the spin rotational symmetry and the point group symmetry. We note

that as an on-site unitary $\mathbb{Z}_2$ symmetry, it is not easy to obtain the fractional quantum number of $(e^{i\pi S^y})^2 = \pm 1$ directly from ground state wave functions [11, 12]. Hence here we take an indirect approach and calculate this fractional quantum number using algebraic relations for the fractionalization of $e^{i\pi S^y}$ and the other point group operations for visons. For simplicity, we use $\omega_X$ to denote fractional quantum numbers: $\omega_{sy} = (e^{i\pi S^y})^2$, $\omega_\mu = \mu^2$, $\omega_\sigma = \sigma^2$, $\omega_I = I_p^2$, $\omega_{sy\mu} = (e^{i\pi S^y}\mu)^2$, $\omega_{sy\sigma} = (e^{i\pi S^y}\sigma)^2$ and $\omega_{syI} = (e^{i\pi S^y}I_p)^2$, respectively; we use $\sigma_{XY}$ to denote the commutation relation fractionalization between two symmetries $X$ and $Y$: $e^{i\pi S^y}\mu = \sigma_{sy\mu}\mu e^{i\pi S^y}$, $e^{i\pi S^y}\sigma = \sigma_{sy\sigma}\sigma e^{i\pi S^y}$ and $e^{i\pi S^y}I_p = \sigma_{syI}I_p e^{i\pi S^y}$. Using the algebraic relation $(XY)^2 = X^2 Y^2 \sigma_{XY}$, we have

$$\omega_{syX} = \omega_{sy}\omega_X \sigma_{syX}, \quad X = \mu, \sigma, I_p. \tag{2}$$

Furthermore, using the relation $I_p = (\sigma\mu)^3$ we derive the following relation,

$$\sigma_{syI} = \sigma_{sy\mu}\sigma_{sy\sigma}. \tag{3}$$

Next, we fix the fractional quantum number of crystal symmetries $\omega_\mu$, $\omega_\sigma$, $\omega_I$, $\omega_{sy\mu}$, $\omega_{sy\sigma}$ and $\omega_{syI}$ using the symmetry eigenvalues of two-vison wave functions. In this exactly solvable model visons are located in the triangles of the kagome lattice, and a wave function containing two visons located in two triangles $i$ and $j$ is obtained by applying the following string operator on the ground state [10],

$$v_{ij} = \pm \prod_{\mathbf{r} \in p(ij)} 2S^z_\mathbf{r}, \tag{4}$$

where the product is taken along a path $p(ij)$ connecting triangles $i$ to $j$, as shown in Fig. 1. The path can only go straight or make "$\pm 30°$" turns, and the $\pm$ sign in the front represents a $\mathbb{Z}_2$ gauge choice depending on the path $p$: if two paths enclose $N$ hexagons, the corresponding string operators differ by a sign $(-1)^N$ because of the constraint that there are three up spins and three down spins on each hexagon [see Fig. 1(c)]. In this model there are two types of visons, denoted as red and blue visons in Ref. 10, located in the up triangles and down triangles of the kagome lattice, respectively. It is easy to check that a string connecting two visons with the same(opposite) colors always contains an even(odd) number of $S^z$.

We begin with the mirror reflection $\sigma$ and consider a string operator along the path $p(ij)$ connecting two triangles $j = \sigma(i)$, as shown in Fig. 1(a). The string operator itself is

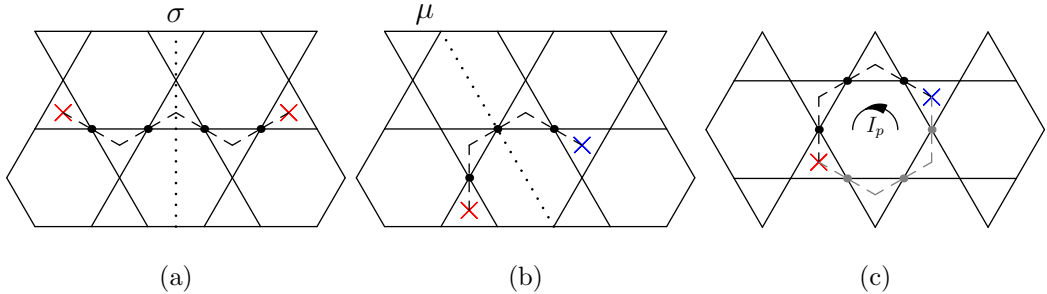

FIG. 1. String operator creating two visons. The dashed line shows a path $p(ij)$ connecting two visons, marked by cross symbols in two triangles of the kagome lattice, where the red and blue color labels visons in up and down triangles, respectively. The string operator is a product of $S^z$ operators on the sites marked by black dots, as in Eq. (4). (a) A vison string operator symmetric under mirror reflection $\sigma$. (b) A vison string symmetric under mirror reflection $\mu$. (c) A string operator connecting two visons symmetric under the 180-degree rotation $I_p$. After $I_p$ the string in black color is mapped to another string in grey color, and the two strings differ by a factor $-1$ because together they enclose one hexagon.

symmetric under $\sigma$, implying that $\omega_\sigma = +1$. Because $\sigma$ maps up triangles to up triangles, the string operator connects two visons with the same color and contains even number of $S^z$, and it commutes with $e^{i\pi S^y}$ [see Fig. 1(a)]. Therefore the two-vison wave function is symmetric under $e^{i\pi S^y}\sigma$, implying that $\omega_{sy\sigma} = +1$.

Similarly for the mirror reflection $\mu$, we consider a string operator along $p(ij)$ where $j = \mu(i)$, as shown in Fig. 1(b). The string operator is still symmetric under $\mu$, hence $\omega_\mu = +1$. However since $\mu$ maps an up triangle to a down triangle and vice versa, the string operator connects two visons with opposite colors and contains an odd number of $S^z$ and therefore anticommutes with $e^{i\pi S^y}$. This implies that $\omega_{sy\mu} = -1$.

Lastly we consider the inversion symmetry $I_p$. In this case the path $p(ij)$ is mapped to another path $I[p(ij)]$, and together with $p(ij)$ it forms an inversion-symmetric loop that encloses an odd number of hexagons because $I_p$ is centered at a hexagon, as shown in Fig. 1(c). Hence the two paths differ by a minus sign, implying that the two-vison wave function is antisymmetric under $I_p$ and therefore $\omega_I = -1$. Furthermore the length of this string is also odd because it connects a up triangle and a down triangle, so the wave function is also antisymmetric under $e^{i\pi S^y}$. Combining these two results we see that the two-vison wave function is symmetric under $e^{i\pi S^y}I_p$ and therefore $\omega_{syI} = +1$.

So far using the vison string operators we explicitly derive the following symmetry fractionalization $\omega_\mu = \omega_\sigma = +1$, $\omega_{sy\mu} = -1$, $\omega_{sy\sigma} = +1$, $\omega_I = -1$ and $\omega_{syI} = +1$. We note that the results of $\omega_\mu$, $\omega_\sigma$, $\omega_I$ and $\omega_{syI}$ are consistent with the results we obtained using general arguments in the main text. Now substituting the results of $\omega_\mu$, $\omega_\sigma$, $\omega_{sy\mu}$ and $\omega_{sy\sigma}$ into Eq. (2), we get

$$\omega_{sy}\sigma_{sy\mu} = -1, \quad \omega_{sy}\sigma_{sy\sigma} = +1. \tag{5}$$

Multiplying these two equations together, we see that $\omega_{sy}$ is canceled and using Eq. (3) we get $\sigma_{sy\mu}\sigma_{sy\sigma} = \sigma_{syI} = -1$. Finally putting this result and $\omega_I = -1$, $\omega_{syI} = +1$ into Eq. (2) we get $\omega_{sy} = +1$.

These results can be understood intuitively by associating the color of vison with its symmetry quantum number $e^{i\pi S^y} = \pm 1$. Since $\omega_{sy} = +1$, the visons carry linear representations of $e^{i\pi S^y}$ and the vison on each triangle carries one of the two possibilities $e^{i\pi S^y} = \pm 1$. Then we consider the string operator in Eq. (4) that creates two visons. Because an even-length string creates two visons with the same color, and the even-length string is invariant under the global symmetry transformation $e^{i\pi S^y}$, two visons with the same color have the same quantum number of $e^{i\pi S^y}$. Similarly as an odd-length string creates two visons with opposite colors and it changes sign under $e^{i\pi S^y}$, the red visons and the blue visons have opposite $e^{i\pi S^y}$ quantum numbers. Without losing generality we assume a red (blue) vison carries $e^{i\pi S^y} = +1 \, (-1)$, respectively. (the two choices are related by a $\mathbb{Z}_2$ gauge transformation and thus physically indistinguishable). This assignment of quantum number is consistent with the commutation relation fractionalization $\sigma_{syI} = -1$, since $I_p$ maps a red vison to a blue vison and therefore anticommutes with $e^{i\pi S^y}$. This also implies that $\sigma_{sy\mu} = -1$ and $\sigma_{sy\sigma} = +1$, because $\mu$ flips vison's color and $\sigma$ preserves vison's color. These are consistent with what we can get from Eq. (5) and $\omega_{sy} = +1$.

In summary, we have shown that the visons in the $\mathbb{Z}_2$ topological state of the BFG model [10] have symmetry fractionalization $(e^{i\pi S^y})^2 = +1$ and $e^{i\pi S^y}$ anticommuting with $I_p$. This explains how the result of $(I_p e^{i\pi S^y})^2 = +1$ from the anomaly test is consistent with $I_p^2 = -1$.

Next we consider symmetry fractionalization involving the time-reversal symmetry. We first notice that when the BFG model is mapped to a quantum dimer model (QDM), the time-reversal symmetry $T$ in the spin model is mapped to a rather unconventional time-reversal symmetry in the dimer model(i.e. it flips the dimer occupations). In fact, $\tilde{T} =$

$Te^{i\pi S^y}$, which is just the complex conjugation $K$, is the more natural time-reversal operation in the dimer model which do not change the dimer configurations. Using this correspondence the symmetry fractionalization in BFG model can be derived from that of the QDM.

First, it is well-known that in the QDM the vison has $\tilde{T}^2 = +1$, thus in the BFG model we get $(Te^{i\pi S^y})^2 = +1$. Furthermore we have $T^2 = +1$ from the flux-fusion anomaly test, and $(e^{i\pi S^y})^2 = +1$ from the previous discussions, so we conclude that $T$ commutes with $e^{i\pi S^y}$.

Second, in the QDM the vison takes trivial symmetry fractionalization $(\mu\tilde{T})^2 = +1$ and $(\sigma\tilde{T})^2 = +1$, which map back to $(\mu Te^{i\pi S^y})^2 = +1$ and $(\sigma Te^{i\pi S^y})^2 = +1$ in the BFG model, consistent with the results of the generalized flux-fusion anomaly test used in the main text. On the other hand, since $e^{i\pi S^y}$ commutes with $T$ and $\sigma$ but anticommutes with $\mu$, we get that $e^{i\pi S^y}$ commutes with $\sigma T$ but anticommutes with $\mu T$. So we find $(\sigma T)^2 = +1$ but $(\mu T)^2 = -1$.

## IV. SYMMETRY FRACTIONALIZATION ON A SQUARE LATTICE

In this section we discuss the symmetry fractionalization for visons on a square lattice, whose space group is $p4mm$. We will also assume there is a half-integer spin per unit cell, and call the bosonic spinon $e$. The algebraic relations and the corresponding invariants are defined in Table I (here we largely follow the notations in Ref. 9 and 13, but we use bond-centered mirror reflection $P_x$ and plaquette-centered inversion $I = P_x P_{xy}$ to ensure that in the flux-fusion anomaly test a symmetric branch-cut can be drawn without going through a site, as discussed in Sec. II).

With U(1)$_{S_z}$ symmetry, we can immediately fix

$$\sigma_{p_x} = \sigma_{p_{xy}} = \sigma_T = 1, \sigma_{t_x t_y} = -1. \tag{6}$$

It is shown in Ref. [14] that $\sigma_{t_x p_x} = 1$.

We now consider the other five invariants, $\sigma_I, \sigma_{t_y p_x}, \sigma_{T p_x}, \sigma_{T p_{xy}}, \sigma_{T t_x}$. To fix those we need to consider SO(3) spin symmetry. Using the same argument as in the main text, we fix $\sigma_{T p_x} = \sigma_I = 1$.

To fix $\sigma_{T t_y}$ (which is equal to $\sigma_{T t_x}$), we put the system on a cylinder geometry where the $y$ direction is compactified, and assume the circumference $L_y \equiv 2 \pmod 4$. We will consider

TABLE I. Algebraic relations on square lattice. The first seven can be constrained by the same argument for the case of triangular lattice. The last three are only present in square lattice.

| Algebraic relation | Invariants for visons | Required symmetry |
| --- | --- | --- |
| $T_x T_y T_x^{-1} T_y^{-1} = \pm 1$ | $\sigma_{t_x t_y}$ | U(1) |
| $P_x^2 = \pm 1$ | $\sigma_{p_x}$ | U(1) |
| $P_{xy}^2 = \pm 1$ | $\sigma_{p_{xy}}$ | U(1) |
| $(P_x P_{xy})^4 = \pm 1$ | $\sigma_I$ | SO(3) |
| $T^2 = \pm 1$ | $\sigma_T$ | U(1) |
| $T P_x T^{-1} P_x = \pm 1$ | $\sigma_{Tp_x}$ | SO(3) |
| $T P_{xy} T^{-1} P_{xy} = \pm 1$ | $\sigma_{Tp_{xy}}$ | SO(3) |
| $T_x P_x T_x P_x^{-1} = \pm 1$ | $\sigma_{t_x p_x}$ | U(1) |
| $T_x P_y T_x^{-1} P_y^{-1} = \pm 1$ | $\sigma_{t_x p_y}$ | U(1) ⋊ $\mathbb{Z}_2$ |
| $T T_x T^{-1} T_x^{-1} = \pm 1$ | $\sigma_{Tt_x}$ | U(1) |

the invariant

$$(TT_y^{L_y/2})^2 = T^2 \cdot (T_y^{L_y/2})^2 \cdot TT_y^{L_y/2} T^{-1} (T_y^{L_y/2})^{-1}, \tag{7}$$

All the quantities are computed on Schmidt states for an entanglement cut along $y$. We view this result as an invariant of the dimensionally reduced 1D SPT, where now $T_y$ is viewed as an on-site symmetry in the 1D system. Using the flux-fusion anomaly test, we can easily prove that $(TT_y^{L_y/2})^2 = 1$ for the vison sector. We have already known that $T^2 = 1$ for the vison. $(T_y^{L_y/2})^2$ is related to the momentum polarization, which is then determined by the topological twist factor of the anyon sector. In general $(T_y^{L_y/2})^2$ is not a 1D SPT invariant, but as shown in Ref. [12], in the presence of $P_y$ it is quantized exactly to the topological twist factor (as long as the magnitude does not vanish), which is 1 for the vison. So we conclude that $TT_y^{L_y/2} T^{-1} (T_y^{L_y/2})^{-1} = \sigma_{Tt_y}^{L_y/2} = 1$ for the vison sector. Since we have chosen a cylinder with $L_y/2$ being an odd integer, we have shown that $\sigma_{Tt_y} = 1$.

A similar argument works for $\sigma_{t_x p_y}$. We still assume $L_y$ is even and consider the translation along $x$. Define $\sigma_{t_x \tilde{p}_y}$ as the ratio of the eigenvalues of $\tilde{P}_y = P_y e^{i\pi S^y}$ per unit length in the $|m\rangle$ and $|\mathbb{1}\rangle$ sectors. Aagain U(1)$_{S_z}$ flux threading fixes $\sigma_{t_x \tilde{p}_y} = 1$, and since we assume $L_y$ is even, the eigenvalue of $e^{i\pi S^y}$ per unit length is 1. We therefore conclude that $\sigma_{t_x P_y} = 1$. Notice that in this argument we do not need to use the connectedness of SO(3), so it would

work equally well if we break the spin rotational symmetry down to U(1) ⋊ $\mathbb{Z}_2$.

In summary, with SO(3) spin rotational symmetry and a half-integer spin per site the symmetry fractionalizations of visons are completely fixed: except $\sigma_{t_x t_y} = -1$, all others must be trivial.

[1] Here $I_p$ includes any inversion that is not centered at a lattice site, including the bond-centered inversion on square and triangular lattices.

[2] M. Cheng, M. P. Zaletel, M. Barkeshli, P. Bonderson, and A. Vishwanath, To appear.

[3] R. Moessner and S. L. Sondhi, Phys. Rev. B **63**, 224401 (2001).

[4] Y. Huh, M. Punk, and S. Sachdev, Phys. Rev. B **84**, 094419 (2011).

[5] Y.-M. Lu, G. Y. Cho, and A. Vishwanath, arXiv:1403.0575 [cond-mat.str-el].

[6] W. Zheng, J.-W. Mei, and Y. Qi, arXiv:1505.05351 [cond-mat.str-el].

[7] Y.-M. Lu, arXiv:1505.06495 [cond-mat.str-el].

[8] X. Yang and F. Wang, arXiv:1507.07621 [cond-mat.str-el].

[9] A. M. Essin and M. Hermele, Phys. Rev. B **87**, 104406 (2013).

[10] L. Balents, M. P. A. Fisher, and S. M. Girvin, Phys. Rev. B **65**, 224412 (2002).

[11] Y. Qi and L. Fu, Phys. Rev. B **91**, 100401(R) (2015).

[12] M. Zaletel, Y.-M. Lu, and A. Vishwanath, arXiv:1501.01395 [cond-mat.str-el].

[13] X.-G. Wen, Phys. Rev. B **65**, 165113 (2002).

[14] M. Hermele and X. Chen, arXiv:1508.00573 [cond-mat.str-el].



\end{document}